%% file: paper.tex
\documentclass[sigplan,10pt]{acmart}
\usepackage{xspace}

\newcommand{\Tn}{{POP}\xspace}
\newcommand{\tn}{{POP}\xspace}
\newcommand{\apn}{{granular}\xspace}
\newcommand{\Apn}{{Granular}\xspace}



\usepackage{algorithm}
\usepackage{algorithmic}
\usepackage{amsmath}
\usepackage[font=small]{caption}
\usepackage{subcaption}
\usepackage[utf8]{inputenc}
\usepackage{booktabs}
\usepackage{tabularx}
\usepackage{grffile}
\usepackage{xcolor}
\usepackage{listings}
\usepackage{hyperref}
\hypersetup{
  colorlinks=true,
  urlcolor=cyan
}

\lstdefinelanguage{Python}{
 keywords={partition, map, reduce},
 keywordstyle=\color{blue}\bfseries,
 basicstyle=\footnotesize\fontencoding{T1}\fontfamily{fvm}\selectfont,
 identifierstyle=\color{black},
 sensitive=false,
 comment=[l]{\#},
 morecomment=[s]{/*}{*/},
 commentstyle=\color{green}\ttfamily,
 string=[s]{"}{"},
 showstringspaces=false,
 stringstyle=\color{violet}\ttfamily,
}
\newcommand{\msr}{{\large$^\dagger$}}
\newcommand{\stanford}{{\large$^\star$}}

\copyrightyear{2021}
\acmYear{2021}
\setcopyright{acmlicensed}\acmConference[SOSP '21]{ACM SIGOPS 28th Symposium on Operating Systems Principles}{October 26--29, 2021}{Virtual Event, Germany}
\acmBooktitle{ACM SIGOPS 28th Symposium on Operating Systems Principles (SOSP '21), October 26--29, 2021, Virtual Event, Germany}
\acmPrice{15.00}
\acmDOI{10.1145/3477132.3483588}
\acmISBN{978-1-4503-8709-5/21/10}

\title{Solving Large-Scale \Apn Resource Allocation Problems Efficiently with POP}
\author{
Deepak Narayanan\stanford, Fiodar Kazhamiaka\stanford, Firas Abuzaid\stanford, Peter Kraft\stanford, Akshay Agrawal\stanford, \\ Srikanth Kandula\msr, Stephen Boyd\stanford, Matei Zaharia\stanford \\
\rm{\textit{\stanford Stanford University\hspace{0.02in} \msr Microsoft Research}}
}

\begin{document}

\input{tex/abstract}

\maketitle

\renewcommand{\shortauthors}{}

\input{tex/abstract}
\input{tex/introduction}
\input{tex/allocation_problems}
\input{tex/overview}
\input{tex/problem_instantiations}
\input{tex/analysis}
\input{tex/implementation}
\input{tex/evaluation}
\input{tex/related_work}
\input{tex/conclusion}
\input{tex/acknowledgements}

\bibliography{paper}
\bibliographystyle{plain}

\newpage
\appendix
\input{tex/appendix}

\end{document}

%% file: tex/abstract.tex
\begin{abstract}

Resource allocation problems in many computer systems can be formulated as
mathematical optimization problems. However, finding exact solutions to these
problems using off-the-shelf solvers is often intractable for large problem
sizes with tight SLAs, leading system designers to rely on cheap, heuristic
algorithms. We observe, however, that many allocation problems are
\emph{granular}: they consist of a large number of clients and resources, each
client requests a small fraction of the total number of resources, and clients
can interchangeably use different resources. For these problems, we propose an
alternative approach that \emph{reuses} the original optimization problem
formulation and leads to better allocations than domain-specific heuristics.
Our technique, Partitioned Optimization Problems (\tn), \emph{randomly} splits
the problem into smaller problems (with a subset of the clients and resources
in the system) and coalesces the resulting sub-allocations into a global
allocation for all clients. We provide theoretical and empirical evidence as to
why random partitioning works well. In our experiments, \tn achieves
allocations within $1.5\%$ of the optimal with orders-of-magnitude improvements
in runtime compared to existing systems for cluster scheduling, traffic
engineering, and load balancing.

\end{abstract}

%% file: tex/introduction.tex
\section{Introduction}

As workloads become more computationally expensive and computer systems become
larger, it has become common for systems to be shared among multiple users. As
a result, deciding how resources (e.g., GPUs, links, servers) should be shared
amongst various clients while optimizing for many macro-objectives is important
across a number of domains (e.g., cluster scheduling, traffic engineering, load
balancing).

\begin{figure}
\center
\includegraphics[width=0.75\columnwidth]{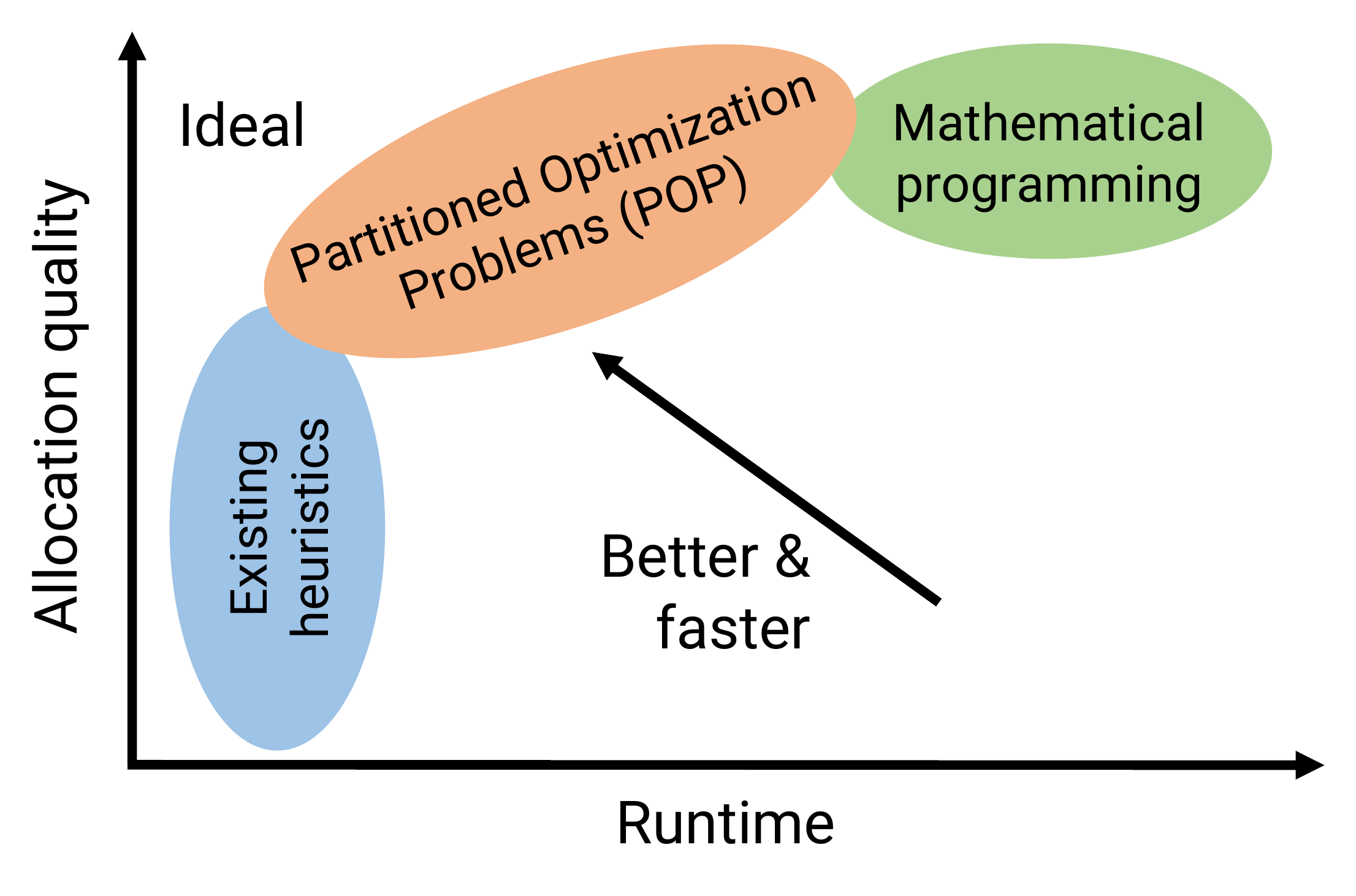}
\vspace{-0.15in}
\caption{
    Tradeoff space between allocation quality (objective-dependent) and
    runtime. Our proposed technique (\tn) is faster than directly solving
    mathematical programs, and computes better allocations than existing
    heuristic algorithms.
    \vspace{-0.1in}
    \label{fig:tradeoff_space}
}
\end{figure}

Resource allocation problems can often be formulated as mathematical
optimization programs~\cite{rizvi2016mayflower, narayanan2020heterogeneity,
taft2014store, li2020traffic, ras, abuzaid2021contracting, pu2015low,
gog2016firmament,kumar2015bwe}; the output of these programs is the allocation
of resources (e.g., accelerators, servers, or network links) to each client
(e.g., jobs, data shards, or traffic commodities). Unfortunately, solving these
mathematical programs can be computationally expensive
(Figure~\ref{fig:tradeoff_space}). The worst-case complexity for linear
programs is approximately
$O(n^{2.373})$~\cite{lee2015efficient,cohen2021solving}, where $n$ is the
number of problem variables (even though LPs can often be solved faster
depending on problem structure), and integer-linear programs are even
more expensive.  Mathematical programs for resource allocation can have
millions of variables (e.g., one variable for every <client, resource> pair)
for large-scale systems, leading to long solution times depending on the numerical
solver used (e.g., 8 minutes for a cluster with 1000 jobs using
SCS~\cite{ocpb:16, scs}). Moreover, allocations often need to be recomputed
frequently to keep up with dynamic changes in the system.
Consequently, production systems such as B4 and
BwE~\cite{hong2018b4,kumar2015bwe} for traffic engineering, the Accordion load balancer~\cite{serafini2014accordion}
for distributed databases, and the Gavel job
scheduler~\cite{narayanan2020heterogeneity}, hit performance bottlenecks when
the numbers of clients and resources increase.

\begin{figure}
    \center
    \includegraphics[width=0.87\columnwidth]{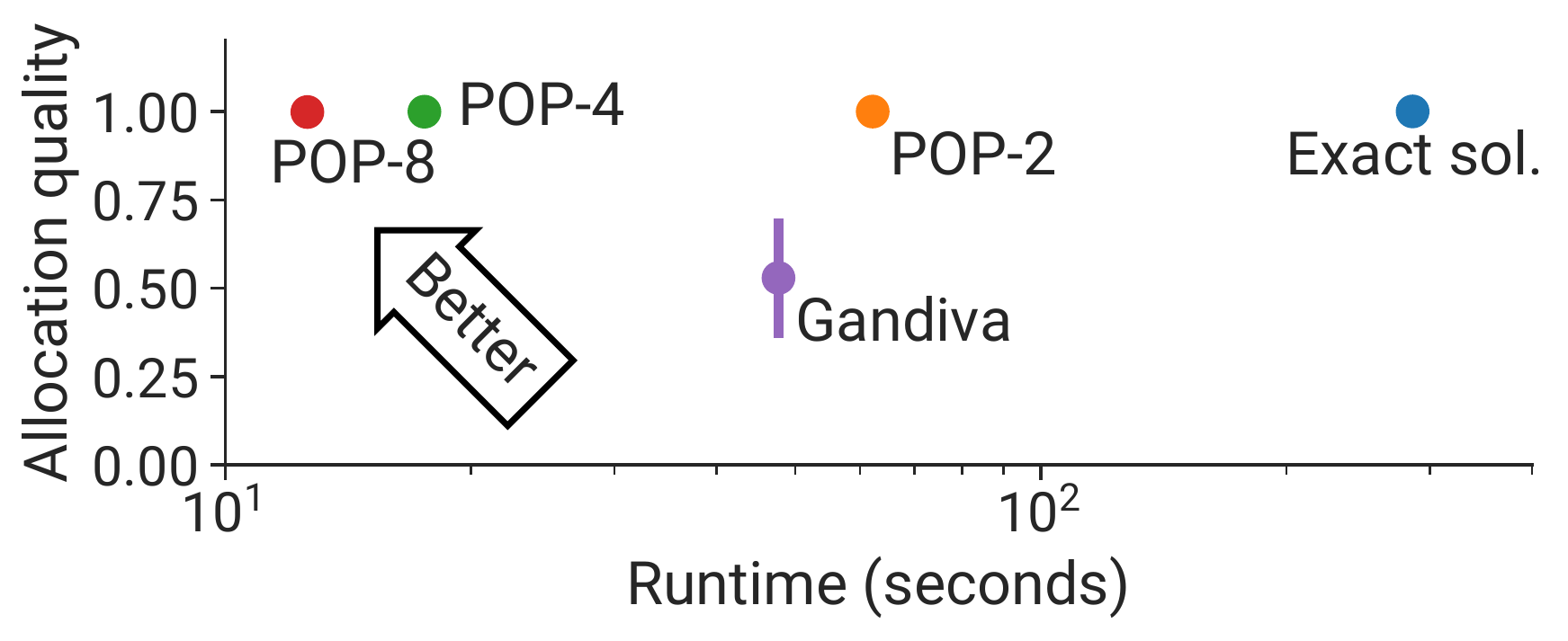}
    \vspace{-0.1in}
    \caption{
        Comparison of Gavel's fair-sharing policy compared to its \tn variants and
        Gandiva~\cite{xiao2018gandiva} on a GPU cluster. The scatterplot shows
        runtimes and mean allocation quality across 2048 jobs on a cluster with
        1536 GPUs. \tn-$k$ uses $k$ sub-problems.
        \label{fig:max_min_fairness_effective_throughput_ratios_and_runtime}
    }
    \vspace{-0.1in}
\end{figure}

Thus, the conventional wisdom in the systems community is that solving these
programs \emph{directly} often takes too long.  Instead, production systems and
researchers frequently use heuristics that are cheaper to compute. It is common
to see some version of the following statement in a paper:
\begin{quote}
``Since these algorithms take a long time, they are not practical for
real-world deployments.  Instead, they provide a baseline with which to compare
faster approximation algorithms.'' -- Taft~\cite{taft2014store}.
\end{quote}
The partition-placement algorithm in E-Store~\cite{taft2014store}, the
space-sharing-aware policy in Gandiva~\cite{xiao2018gandiva}, and cluster
management policies to allocate resources to containers in systems like
Kubernetes~\cite{kubernetes}, DRS~\cite{gulati2012vmware}, and
OpenShift~\cite{openshift} all rely on heuristics. However, prior work shows
that these heuristics are hard to maintain as problems scale and inputs
change~\cite{suresh2020scalable}, are far from optimal
(Figures~\ref{fig:max_min_fairness_effective_throughput_ratios_and_runtime},
\ref{fig:total_flow_and_runtime}, and
\ref{fig:number_of_shard_movements_and_runtime}), and often do not extend to
slightly modified objectives.

Although it seems that large optimization problems are too expensive to solve
directly, we observe that many allocation problems in computer systems share
several exploitable properties: the number of clients and resources is large,
each client requests a small fraction of the total number of resources, and
resources are fungible or substitutable (i.e., a job can make similar progress
using \emph{different} resources). For any such \emph{granular} allocation
problem, we propose \tn, which stands for Partitioned Optimization Problems: a
method for quickly computing allocations by \emph{reusing} the original
optimization problem formulation on subsets of the input. On several granular
optimization problems, we found that \tn can give close-to-optimal results with
orders-of-magnitude faster runtimes than the full formulations. Importantly,
since \tn reuses the original problem formulations, it can be implemented in
only a few lines of code.

The simplest way to apply \tn is to divide clients and resources among $k$
identical copies of the given optimization problem (each with a subset of the
clients and resources). Each sub-problem has fewer equations and variables,
leading to a super-linear runtime speedup. The sub-problems can also be
executed in parallel. The overall allocation is a union of the allocations from
the individual sub-problems. Our results show that randomly and evenly dividing
clients and resources among sub-problems works well when clients are numerous
and individually use only a small fraction of all resources. Empirically, we
show that \tn's resource allocations are nearly optimal on several
optimization problems, including using real-world inputs. We also prove that
the probability of a large optimality gap is small given an allocation problem
with certain simple properties. \tn has structural similarity with the first
step of ``primal decomposition'' in convex optimization~\cite{boyd2007notes},
but can be applied to a broader set of problems than those amenable to primal
decomposition (separable objective, coupled constraints). We
note that there can be other ways to \tn an optimization problem, but these are
beyond the scope of this paper.

In the wild, allocation problems do not always fit the definition of granular
as presented above, e.g., a traffic engineering problem could have ``large''
clients (commodities) with substantial bandwidth demand, or a client might have
to use a particular resource (e.g., link between two sites). Fortunately, in
some such cases, we can transform the problem into a granular problem using two
\emph{granularization} techniques: \emph{client splitting} and \emph{resource
splitting}.  The ``large'' clients, which individually require a sizable
fraction of total resources, can be split into multiple virtual clients who each
receive partial allocations from multiple sub-problems. Since the number of
``large'' clients is small, by definition, \tn's sub-problems remain small and
still achieve a sizable runtime speedup. Similarly, resources can be split into
multiple virtual resources, each with a fraction of the full resource's
capacity.

\Tn cannot be applied to every allocation problem in systems because some
problems are not granular or require a non-trivial partitioning into
sub-problems (e.g., due to constraints).  We discuss examples of such problems
in \S\ref{sec:non_applicable_problems}.

Nevertheless, we found that \tn is effective on a wide range of important problems
in recent computer systems research. We
evaluated \tn on 6 different allocation objectives across three domains (cluster
scheduling, traffic engineering, and load balancing). \tn achieves empirical
runtime improvements of up to $100\times$ compared to the original optimization
problem formulations while staying within 1.5\% of optimal, and even up to
$20\times$ faster and $1.9\times$ higher quality than heuristics. We integrated
\tn into real systems like Gavel, and found that downstream metrics like
average job completion time and makespan are unaffected by using \tn.  We also
found granularization useful in using \tn to compute high-quality allocations
for initially non-granular problems, like traffic engineering problems with a
few large flows and links between specific sites. Our implementation is
available at \url{https://github.com/stanford-futuredata/POP}.

%% file: tex/allocation_problems.tex
\section{\Apn Allocation Problems}
\label{sec:granular_allocation_problems}

Computer systems are often shared among \emph{clients} from multiple users
(e.g., jobs in a cluster scheduler, commodities in a Wide Area Network). These
clients might then request \emph{resources} (e.g., GPUs or link capacity)
from a central resource allocator, which determines how to map
resources to clients. Resource allocation problems have three main components:
\begin{itemize}

    \item \textbf{Search Space of Allocations:} Allocations specify how
resources should be shared between clients. In cluster scheduling, an
allocation can specify the fraction of wall-clock time each active job should
spend on different types of resources (e.g., types of GPUs like K80, P100,
V100, A100). In traffic engineering, an allocation can specify the flow each
commodity should receive on different links. Allocations can also reason
through the interactions between clients on different resources (e.g., the time
fractions \emph{pairs} of jobs should spend on various
resources~\cite{xiao2018gandiva, narayanan2020heterogeneity}).

    \item \textbf{Objectives:} The objective that an optimization problem
maximizes or minimizes is a function over the allocation, and specifies the
metric (e.g., dollar cost, total flow) that needs to be optimized in solving
the allocation problem. We observe that these functions are typically a $\max$
or sum over functions of per-client allocations, but can be other arbitrary
functions as well. Convex functions are generally easier to optimize.

    \item \textbf{Constraints:} Most allocation problems also specify
constraints to ensure that both clients and resources are not over-allocated
(e.g., the total time fraction given to a single job across resource types
cannot exceed 1.0) and that various invariants are maintained. These are
specified as functions over the allocation $A$.

\end{itemize}

The goal of a resource allocation problem is to find the allocation value that
is feasible (respects the provided constraints) and optimizes the provided
objective.

\vspace{0.1in}
We can then say an allocation problem is \textbf{\apn} if:
\begin{itemize}

    \item \textbf{Condition 1:} The number of clients and resources is large
(on the order of 100s or more).

    \item \textbf{Condition 2:} Each client requests an insignificant fraction
(e.g., $<1\%$) of the total available resources.

    \item \textbf{Condition 3:} Resources are fungible or substitutable.  In
other words, if a client $c$ is given resource $r$ as part of an allocation
$A$, there are multiple other resources $r'\neq r$ such that switching $c$ to
$r'$ gives an allocation $A'$ with similar objective value ($f(A) \approx
f(A')$).
    
    \item \textbf{Condition 4:} If the resource allocation problem considers
interactions between multiple clients (e.g., two jobs on the same server), then
client combinations should be fungible or substitutable too.

\end{itemize}

As we show in \S\ref{sec:prob-instances}, resource allocation problems in a
number of different domains like cluster scheduling, traffic engineering, and
load balancing, are \apn. Furthermore, in certain cases, problems that violate
some of these conditions can be made \apn through granularization
transformations (client and resource splitting in
\S\ref{sec:client_and_resource_splitting}).

For example, in Gavel~\cite{narayanan2020heterogeneity}, a cluster scheduler
for machine learning training workloads on clusters of GPUs, each job (client)
requests a prescribed number of a resource (e.g., a specific kind of GPU) to
make progress. Each job requests a small fraction of the total number of GPUs
available in the cluster, and can be run on different types of GPUs with
varying efficiencies. Additionally, when used with space
sharing~\cite{xiao2018gandiva, narayanan2020heterogeneity}, each job can be run
with many other jobs (again with varying efficiencies).  We assume that
dependencies that specify when jobs are runnable are handled by a separate DAG
scheduler. This is standard in systems such as Spark and
Hadoop~\cite{zaharia2012resilient}. Such cross-job ``when can job $X$ run''
dependencies are not under the purview of the resource schedulers considered in
this paper, which try to determine how resources should be shared among already
\emph{runnable} jobs only.

In traffic engineering setups such as those considered in
NCFlow~\cite{abuzaid2021contracting}, the clients are commodities, each
resource is a network link between two sites in the Wide Area Network, and each
commodity typically requests a small fraction of the total available capacity.

In load balancing, the clients are data shards, the resources are servers, and
each shard can be handled by a small fraction of the total number of servers
available in the system.

%% file: tex/overview.tex
\section{Partitioned Optimization Problems}

\Apn resource allocation problems can be split into sub-problems, where each
sub-problem has a subset of the clients and resources in the full allocation
problem. We leverage the large number of clients and resources to randomly
partition clients and resources into sub-problems; this procedure yields
high-quality allocations due to the law of large numbers. We call this
technique Partitioned Optimization Problems (or \tn for short). In the rest of
this section, we describe the intuition, procedure, and benefits of POP.

\subsection{Intuition}

Optimization problems for large systems take a long time to solve in part
because they have many variables.  For example, consider an optimization
problem that involves scheduling $n$ jobs on $m$ cloud VMs. Each VM has varying
amounts of resources (e.g., CPU cores, GPUs, and RAM). To express the
possibility of any job being assigned to any VM, an $n\times m$ matrix of
variables would be needed; for $10^4$ jobs and $10^4$ VMs, the problem has
$10^8$ variables.  Contemporary solvers often take hours to solve such
problems, although the exact runtime depends on problem properties such as
sparsity \cite{vanderbei2015linear}.

We can achieve much faster allocation computation times by decomposing the
problem; for example, the problem of scheduling $10^3$ jobs on $10^3$ VMs
($100\times$ fewer variables) is much more tractable. This procedure of
breaking up the larger problem into sub-problems \emph{reduces the search
space} explored by the solver, since interactions between all combinations of
clients and resources are no longer considered. Instead, only combinations of
\emph{subsets} of clients and resources are considered, which reduces runtime
but also can reduce the quality of the allocation. In light of this, the
interaction between clients and resources needs to be considered carefully to
take into account the many global constraints in the original problem, as well
as the objective (e.g., fairness). We find that on large granular resource
allocation problems, splitting clients \emph{randomly} and assigning an equal
number of resources among sub-problems reduces the search space of feasible
solutions that needs to be considered by solvers, while still ensuring that
\emph{some} high-quality feasible points are in the explored search space.
This is the main intuition that allows \tn to be effective, returning
allocations of similar quality as the original formulation but faster.

\subsection{Procedure for \Apn Problems}

\begin{algorithm}[tb]
   \caption{\tn Procedure.}
   \label{alg:pop_procedure}
\begin{algorithmic}
   \STATE {\bfseries Input:} Clients and their attributes $X = [x_1, x_2, \ldots, x_n]$, resources and their attributes $Y = [y_1, y_2, \ldots, y_m]$, function to compute allocations $\textsc{get\_allocation}:(X, Y) \rightarrow A$, number of partitions $k$, (optional) splitting attribute $s$, (optional) ratio of extra virtual clients allowed $t$.
   \STATE {\bfseries Return:}  Allocation for all $n$ clients, $A$.
   \STATE
   \STATE \textit{// Optional: make the problem granular if it is not already.}
   \STATE $X' = \textsc{split\_clients}(X, s, t)$, $Y' = \textsc{split\_resources}(Y)$
   \STATE
   \STATE \textit{// This is the \lstinline{partition} step.}
   \STATE $[X'_1, X'_2, \ldots, X'_k], [Y'_1, Y'_2, \ldots, Y'_k] = $ \lstinline{partition}$(X', Y', k)$
   \STATE
   \STATE \textit{// This is the \lstinline{map} step, can be performed in parallel.}
   \FOR{$i$ in \lstinline{range}$(k)$}
   \STATE $A_i = \textsc{get\_allocation}(X'_i, Y'_i)$
   \ENDFOR
   \STATE
   \STATE \textit{// This is the \lstinline{reduce} step; allocations $A_i$ are combined.}
   \STATE $A = \textsc{coalesce}([A_1, A_2, \ldots, A_k])$
\end{algorithmic}
\end{algorithm}

The first step of \tn is to \lstinline{partition} larger allocation problems
into smaller allocation sub-problems. The type of partitioning allowed is
dependent on the objective and constraints of the allocation problem, and has
implications on the runtime speedups and quality of the returned allocation. We
can then re-use the map-reduce API~\cite{dean2008mapreduce,
zaharia2012resilient} (or divide-and-conquer): each of these sub-problems can
be solved in parallel (\lstinline{map} step), and then allocations from the
sub-problems can be reconciled into a larger allocation for the entire problem
(\lstinline{reduce} step). We show pseudocode for this in
Algorithm~\ref{alg:pop_procedure}.

\begin{figure}
    \center
    \vspace{-0.1in}
    \includegraphics[width=0.95\columnwidth]{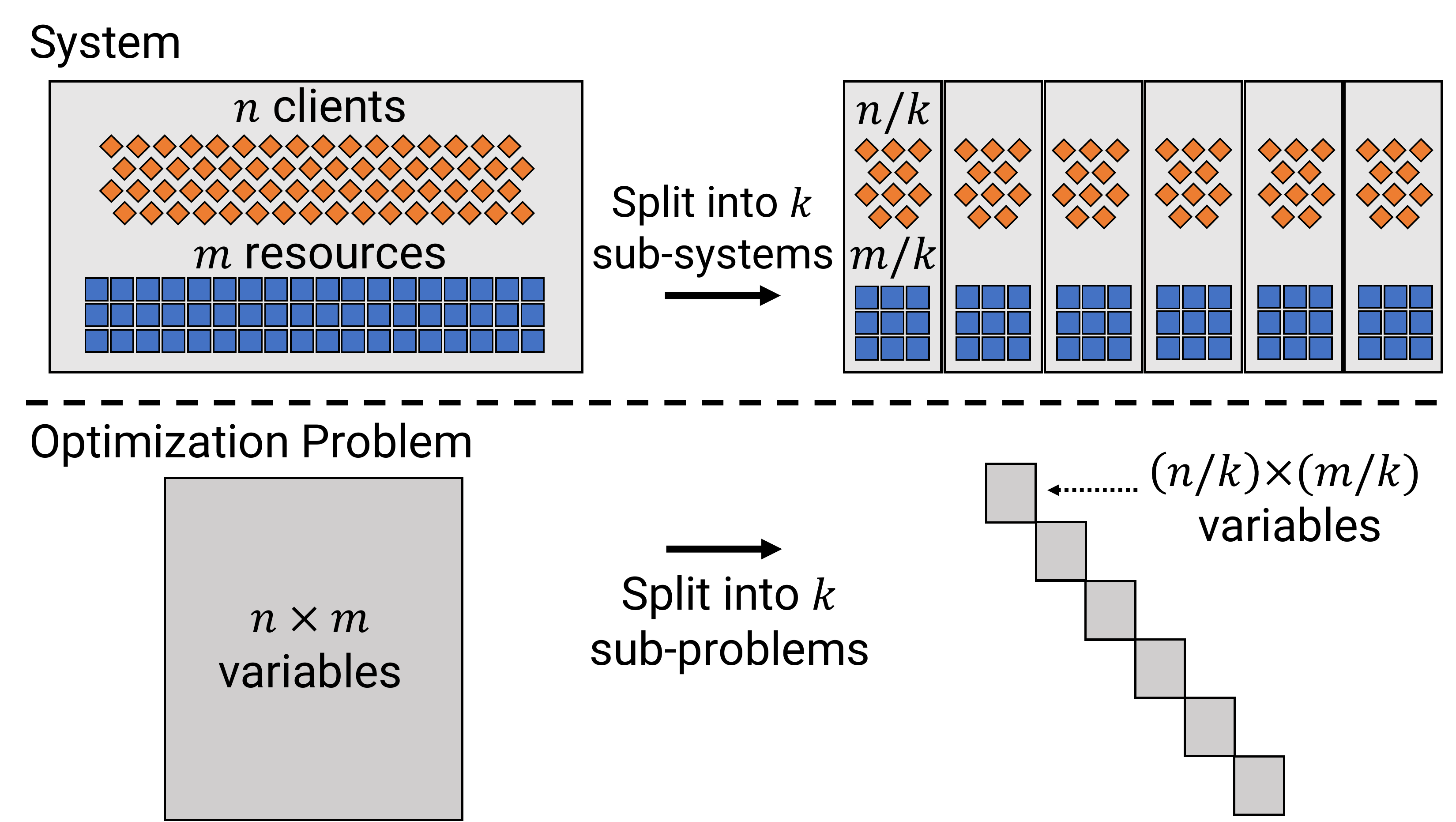}
    \vspace{-0.1in}
    \caption{
        \tn partitions the system to reduce the number of optimization problem variables.
        For a problem where the number of variables is the number of clients times the
        number of resources, dividing clients and resources evenly among $k$ sub-problems
        reduces the number of variables in each sub-problem by $k^2$.
        \label{fig:system_scaling}
    }
    \vspace{-0.1in}
\end{figure}

The partitioning step affects the runtime, the reconciliation complexity, and
ultimately the quality of the final allocation. One straightforward approach
that we explore in this paper is to divide \emph{both} clients (e.g., jobs,
shards, flows) and resources (e.g., servers, links) randomly into
\emph{sub-systems}, as shown in the top half of
Figure~\ref{fig:system_scaling}. We find that this partitioning scheme is
effective even when clients have attributes with skew (e.g., jobs in a shared
cluster with various priority levels, or data shards in query load balancing
with different loads). Low-quality allocations can also result from clients
having vastly different utilities with different resources. For example, a
resource could be a network link between two sites in a Wide Area Network
(WAN). A commodity might \emph{have} to use this link to send traffic between
these two sites.  This paper shows how client and resource splitting
(\S\ref{sec:client_and_resource_splitting}) can be used to transform some of
these ``hard'' problems into a form that is then amenable to \emph{random
partitioning}. Other broad partitioning strategies can also be used depending
on problem structure (e.g., assign all ``geographically close'' clients and
resources to the same sub-problem), but these are out of the scope of this
paper.  With random partitioning, the \lstinline{reduce} step is cheap, as
simply concatenating sub-system allocations yields a feasible allocation to the
original problem.

\subsection{Transformations to Granularize Problems}
\label{sec:client_and_resource_splitting}

In some cases, it might not be possible to either return an allocation that is
feasible or high quality by merely assigning each client and resource to
sub-problems at random when using the \tn procedure. Skewed workloads with
heavy tails are common in practice~\cite{tirmazi2020borg}. As an example,
consider a query load balancing problem where we try to assign shards
containing various keys to compute servers: our goal is to spread load evenly
amongst the available servers, which can be formulated as a mixed-integer
linear program (\S\ref{sec:prob-lb}).  In such a setting, it is common for
single shards to be \emph{hot}: for example, Taylor Swift's Twitter account
receives much more request traffic compared to the average Twitter user. In
light of these hot shards, it might not be possible to assign shards to
individual sub-problems and obtain sub-problems with input distributions
similar to the original problem (and consequently leading to either an
infeasible or poor-quality allocation).  Similarly, in the traffic engineering
problem, it is common for a small number of commodities to have large
demands~\cite{abuzaid2021contracting}. Multiple such commodities in a
sub-problem would lead to sub-optimal total flow.  To transform these into
granular problems, we propose an algorithm to \emph{split} variables for
clients and resources across several sub-problems.

\begin{figure}
    \center
    \includegraphics[width=\columnwidth]{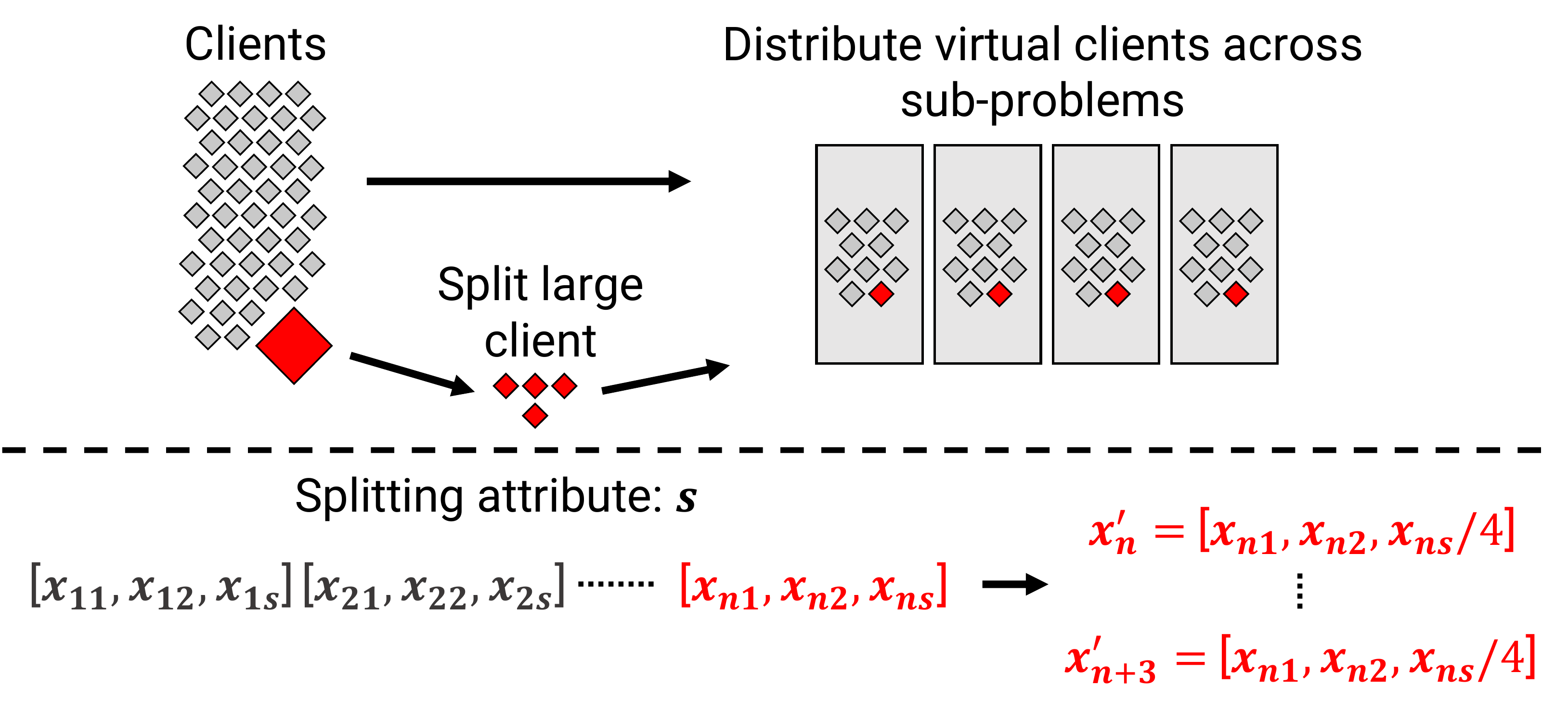}
    \vspace{-0.3in}
    \caption{
        Client splitting, where we \emph{granularize} non-granular
        problems by splitting clients based on a splitting attribute $s$.
        \label{fig:client_splitting}
    }
\end{figure}

\begin{algorithm}[tb]
   \caption{Client Splitting Algorithm.}
   \label{alg:splitting_algorithm}
\begin{algorithmic}
   \STATE {\bfseries Input:} Inputs $X=[x_1, x_2, \ldots, x_n]$, splitting attributes $s$, ratio of extra virtual clients $t$ allowed.
   \STATE {\bfseries Return:} Mapping from real to virtual clients $\{x_i \rightarrow [x'_j]\}$.
   \STATE Initialize $\text{queue} \leftarrow \textsc{max\_heap}()$, $\text{mapping} \leftarrow \{\}$.
   \STATE For all $i \in \{1, 2, \ldots, n\}$, $\text{queue}.\textsc{push}(x_i.s, x_i)$.
   \STATE
   \WHILE{$\text{len}(\text{queue}) \leq (1+t) \cdot n$}
   \STATE $x_\text{max} = \text{queue}.\textsc{pop}()$
   \STATE Split $x_\text{max}$ by attribute $s$ into two copies $x_\text{max}^1$ and $x_\text{max}^2$ 
   ($x_\text{max}^1.s, x_\text{max}^2.s = x_\text{max}.s / 2$).
   \STATE
   \STATE $\textsc{update\_mapping}(x_\text{max}, [x_\text{max}^1, x_\text{max}^2])$
   \STATE $\text{queue}.\textsc{push}(x_\text{max}^1.s, x_\text{max}^1)$, $\text{queue}.\textsc{push}(x_\text{max}^2.s, x_\text{max}^2)$
   \ENDWHILE
\end{algorithmic}
\end{algorithm}

\paragraph{Client Splitting.}

We require the user to specify the client attribute that specifies resource
demand and can be split across several sub-problems; all other attributes are
copied over without change. In the load balancing example, where clients are
data shards and attributes include  shard load and memory size, the
\emph{splitting attribute} is the shard load. In the traffic engineering
example, the splitting attribute is the commodity's traffic demand. Given this
splitting attribute, we then construct a priority queue (heap) of the
corresponding attribute values for all clients. Given a threshold $t$ ($t$
is typically a number less than 1) on the
maximum number of extra \emph{virtual} clients allowed, we pop and split
variables off the queue, and then push the new variables back into the queue.
Each split reduces the value of the splitting attribute of the popped variable
by a factor of 2. Importantly, each split maintains the feasibility invariant:
the coalesced allocation across virtual clients will still be feasible (since
the total sum of splitting attribute values remains the same). By reducing the
value of the splitting attribute, client splitting breaks down large clients
into a collection of smaller clients with \emph{equivalent} total demand. The
runtime of this algorithm is $O(n \log n)$, where $n$ is the number of clients,
which is cheap compared to the runtime of allocation computation in each
sub-problem.  Algorithm~\ref{alg:splitting_algorithm} shows pseudocode, and the
procedure is illustrated in Figure~\ref{fig:client_splitting}. Empirically, we
found that most problems are granular enough for POP to work well with 0 split
clients. Client splitting does not adversely impact allocation quality, but can
increase runtime. The hardest problems in our experiments required $t=0.75$.
The optimal value of $t$ is problem-specific and it is possible that users may have to dynamically adapt $t$ to get the best performance from \tn; however, in all of the considered production use-cases in our experiments, we found that small values of $t$ that worked well for historical problem instances continue to work well on future problem instances.

\paragraph{Resource Splitting.}

If a client has to use a particular resource to make progress, \tn will not
work out of the box, since randomly partitioning clients and resources into
sub-problems might result in a partitioning where the client is not matched
with its preferred resource. In such cases, each resource can be split into $k$
``virtual'' resources (where $k$ is the number of sub-problems). Each virtual
resource has $k\times$ lower capacity, and is assigned to a different
sub-problem. By ensuring that each virtual resource has lower capacity, we
ensure that the final coalesced allocation is still feasible.

\vspace{0.15in}

Client and resource splitting are not always applicable. For example, resource
splitting cannot be used easily if the allocation problem's objective depends on
whether a resource is used or not (e.g., an allocation problem that tries to
minimize the number of resources used). Similarly, client splitting cannot be
used easily for problems which take into account interactions between multiple
clients sharing a resource.

The resulting allocation problem after these transformation steps can be \apn;
if so, we can use \tn to solve it. After the \lstinline{partition} step, we
obtain allocations for each virtual variable in the problem. Allocations
assigned to virtual variables corresponding to a single client need to be
summed to obtain the final allocation. We show how this can be incorporated
into the full \tn procedure in Algorithm~\ref{alg:pop_procedure}.

\subsection{Benefits of \tn}
\Tn has several desirable properties:

\begin{itemize}

    \item \textbf{Simplicity:} Users do not need to design new heuristics from
scratch to scale up to larger problem sizes, and can reuse their original
problem formulations.

    \item \textbf{Generality across domains and solvers:} \Tn can be used to
accelerate allocation computations for many different types of problem
formulations across domains. \Tn also easily integrates with different
solvers.

    \item \textbf{Applicability to different types of objectives:} \Tn can be
applied for a broad class of objectives, such as total flow and maximum
concurrent flow in traffic engineering. These objectives have traditionally
required very different approximation algorithms~\cite{karakostas2008faster,
fleischer2000approximating}.

    \item \textbf{Composability:} \Tn can be used for any granular allocation
problem in an outer loop as a simplifying step; existing heuristics or
approximation algorithms can then be used to solve the resulting sub-problems.

    \item \textbf{Tunability:} The number of sub-problems is a knob for trading
off between allocation quality and runtime.
\end{itemize}

%% file: tex/problem_instantiations.tex
\section{Case Studies of Applying \Tn} \label{sec:prob-instances}

In this section, we describe various resource allocation problems that are
formulated as optimization problems: scheduling of jobs on clusters with possibly
heterogeneous resources~\cite{narayanan2020heterogeneity}, WAN traffic
engineering~\cite{abuzaid2021contracting}, and query load
balancing~\cite{serafini2014accordion, taft2014store, curino2011workload}.  We
show the full \emph{exact} problem formulations presented in the corresponding
papers, and then explain how \tn can be used to compute high-quality
allocations faster. We also present some examples of problems which are not
\apn and out of scope for \tn.

\subsection{Resource Allocation for Heterogeneous Clusters} \label{sec:prob-cs}

We first discuss the optimization problem formulations used in Gavel, which
supports a range of complex objectives. These can be accelerated
using \tn since these problems are granular, i.e., meet the
conditions in \S\ref{sec:granular_allocation_problems}.

Gavel~\cite{narayanan2020heterogeneity} is a cluster scheduler that assigns
cluster resources to jobs while optimizing various multi-job objectives (e.g.,
fairness, makespan, cost). Gavel assumes that jobs can be time sliced onto the
available heterogeneous resources, and decides what fractions of time each job
should spend on each resource type by solving an optimization problem.
Optimizing these objectives can be computationally expensive when scaled to
1000s of jobs, especially with ``space sharing'' (jobs execute concurrently on
the same resource), which requires variables for every \emph{pair} of runnable
jobs.

Allocation problems in Gavel are expressed as optimization problems in terms of
a quantity called \emph{effective throughput}: the throughput a job observes
when given a resource mix according to an allocation $A$, computed as:
$$\text{throughput}(\text{job } j, \text{allocation }A) = \sum_i T_{ji} \cdot
A_{ji}.$$ $T_{ji}$ is the raw throughput of job $j$ on resource type $i$. In
Gavel, vanilla heterogeneity-aware allocations $A_{ji}$ are assigned to each
combination of job $j$ and GPU type $i$. $A_{ji}$ represents the fraction of
wall-clock time that a job $j$ should spend on the GPU type $i$.  We now show
formulations for three objectives.

\paragraph{Max-Min Fairness.} The Least Attained Service
policy~\cite{gu2019tiresias} tries to give each job an equal resource share of
the cluster. The heterogeneity-aware version of this policy can be expressed as
a max-min optimization problem over all active jobs in the cluster. We assume
that each job $j$ has fair-share weight $w_j$ and requests $z_j$ GPUs. Then, to
take into account the impact of moving a job between GPU types, we find the
max-min allocation of normalized effective throughputs:
$$\text{Maximize}_A \min_j \dfrac{1}{w_j} \dfrac{\text{throughput}(j, A)}{\text{throughput}(j, A^{\text{equal}})} \cdot z_j.$$

$A^\text{equal}$ is the allocation given to job $j$ assuming it receives equal
time share on each worker type in the cluster. We also need to specify
constraints to ensure that jobs and the cluster are not over-provisioned
(e.g., total GPU allocation time does not exceed the total number of GPUs):
\begin{eqnarray}
& 0 \leq A_{ji} \leq 1 & \forall (j, i) \nonumber \\
& \sum_i A_{ji} \leq 1 & \forall j \nonumber \\
& \sum_j A_{ji} \cdot z_j \leq \text{num\_workers}_i & \forall i \nonumber
\end{eqnarray}

The above formulation can be extended to consider space
sharing~\cite{xiao2018gandiva, narayanan2020heterogeneity}, where multiple jobs
execute concurrently on the GPU to improve GPU utilization, by only changing the
way effective throughput is computed; see the Gavel
paper~\cite{narayanan2020heterogeneity} for details.

\paragraph{Proportional Fairness.} Proportional
fairness~\cite{agrawal2021allocation} tries to maximize total utilization while
still maintaining some minimum level of service for each user (in this case,
job).  Proportional fairness for GPU cluster scheduling can be formulated as
the following convex optimization problem: $$\text{Maximize}_A \sum_j \log
(\text{throughput}(j, A)).$$ Constraints are the same as before. Per-job
weights and other extensions are also possible (the above objective can be
interpreted as a sum of utilities, i.e., $\text{Maximize}_A \sum_i U_i(A_i)$).

\paragraph{Minimize Makespan.} We can also minimize makespan (the time taken by
a collection of jobs to complete) using a similar optimization problem
framework. Let $\text{num\_steps}_j$ be the number of iterations remaining to
train job $j$.  The makespan can then be computed as the maximum of the
durations of all active jobs; the duration of job $j$ is just the ratio of the
number of iterations to $\text{throughput}(j, A)$. Mathematically, this can be
written as follows using the same above constraints:
$$\text{Minimize}_A \max_j \dfrac{\text{num\_steps}_j}{\text{throughput}(j, A)}.$$

\paragraph{Using \tn.} We can use POP on these cluster scheduling problems by
partitioning the full set of jobs into job subsets, and the cluster into
sub-clusters. Each sub-cluster has an equal number of resources (GPUs of each
type), and jobs are partitioned randomly into the job subsets. The \tn solution
is feasible by construction. Since the cluster has multiple resources of each
type (e.g., GPU of specific generation), the problem is \apn by default, and
does not require additional transformations to be made \apn.  Additionally,
even when allowing job colocation (using space sharing), jobs can make progress
colocated with many other jobs.

\subsection{Traffic Engineering and Link Allocation} \label{sec:prob-te}

We next discuss optimization problem formulations that require both resource and
client splitting to be solved accurately and efficiently by \tn.

The problem of traffic engineering for networks determines how flows in a Wide
Area Network (WAN) should be allocated fractions of links of different
capacities to best satisfy a set of demands.  One might consider several
objectives, such as maximizing the total amount of satisfied flow, or
minimizing the extent to which any link is loaded to reserve capacity for
demand spikes.

\paragraph{Maximize Total Flow.} The problem of maximizing the total flow,
given a matrix of per-commodity demands $D$ (each commodity or flow $j$ has a
demand $D_j$), a pre-configured set of paths $P$, and a list of edge capacities
$c_e$, can be written as:
$$\text{Maximize}_A \sum_{j \in D} A_j.$$
Subject to the constraints:
\begin{eqnarray}
& A_j = \sum_p A_j^p  & \forall j \in D \nonumber \\
& A_j \leq D_j        & \forall j \in D \nonumber \\
& \sum_{\forall j, p \in P_j, e \in p} A_j^p \leq c_e & \forall e \in E \nonumber \\
& A_j^p \geq 0 & \forall p \in P, j \in D \nonumber
\end{eqnarray}

$A_j^p$ is the flow assigned to commodity $j$ along path $p$ (one of the paths
in $P_j$). The constraints ensure that the total flow through an edge does not
exceed the capacity of the edge, that each commodity's flow per path is
positive, and each commodity's flow does not exceed its demand.

For every commodity, the set $P$ consists of a pre-computed set of
paths between the source and target nodes~\cite{abuzaid2021contracting}.

\paragraph{Maximize Concurrent Flow.} The objective only needs to be changed to:
$$\text{Maximize}_A \min_{j \in D} A_j.$$
The constraints are the same as above.

\paragraph{Using \tn.} To accelerate allocation computation using \tn, we need
to granularize the original problems. In particular, we use resource splitting
for all traffic engineering problems: we assign the entire network (all nodes
and edges) to each sub-problem (but each link with a fraction of the total
capacity), and distribute commodities across sub-problems. We do not shard the
network itself (i.e., assign each link to a single sub-problem only) since
traffic can flow between any pair of nodes and the difference in utility for
any commodity when using a fraction of the available links in the network is
high (links between specific sites may \emph{need} to be used to sustain
sufficiently high flow). By assigning each sub-problem a link with a fraction
of the total capacity, we ensure that the final allocation from \tn is
feasible. For specific problems with large commodities, we also use client
splitting.

\subsection{Query Load Balancing} \label{sec:prob-lb}

Systems like Accordion~\cite{serafini2014accordion},
E-Store~\cite{taft2014store}, and Kairos~\cite{curino2011workload} need to
determine how to place data items in a distributed store to spread load across
available servers.

We consider the problem of load balancing data shards (collections of data
items).  This is similar to the single-tier load balancer in E-Store, but
acting on collections of data items instead of individual tuples.  The
objective is to minimize shard movement across servers as load changes, while
constraining the load on each server to be within a tolerance $\epsilon$ of
average system load $L$.  Each shard $i$ has load $l_i$ and memory footprint
$f_i$. Each server $j$ has a memory capacity of $\text{memory}_j$ that
restricts the number of shards it can host.  The initial placement of shards is
given by a matrix $T$, where $T_{ij}=1$ if partition $i$ is on server $j$. $A$
is a shard-to-server map, where $A_{ij}$ is the fraction of queries on
partition $i$ served by $j$, and $A'_{ij} = 1$ if $A_{ij} > 0$, 0 otherwise.
Finding the balanced shard-to-server map that minimizes data movement can then
be formulated as a mixed-integer linear program:
$$\text{Minimize}_{A} \sum_{i} \sum_{j} (1 - T_{ij}) A'_{ij} f_i.$$
Subject to the constraints:
\begin{eqnarray}
& L - \epsilon \le \sum_i A_{ij}l_i \le L + \epsilon & \forall j \nonumber \\
& \sum_j A_{ij} = 1 & \forall i\nonumber \\
& \sum_i A'_{ij}f_i \le \text{memory}_j & \forall j \nonumber \\
& A_{ij} < A'_{ij} \leq A_{ij} + 1 & \forall (i,j) \nonumber
\end{eqnarray}

\paragraph{Using \tn.} The load balancing problem can be accelerated using \tn
by dividing the shard set and server cluster into shard subsets and server
sub-clusters, while ensuring that each shard subset has the same total load.

\subsection{When is \Tn Not Applicable?}
\label{sec:non_applicable_problems}

Although \tn can be used on a number of different resource allocation problems,
it cannot be used for \emph{all} possible problem formulations. Here, we
present a few examples of resource allocation problems where \tn with random
partitioning \emph{cannot} be used.

\paragraph{Capacitated Facility Location.} The capacitated facility location
problem tries to minimize the cost of satisfying users' demand given a set of
processing facilities. Each facility has a processing capacity, and also a
``leasing cost'' if used at all (if a facility is not processing any demand, it
has a leasing cost of 0). The cost of processing some demand by a facility is
proportional to the distance of the facility from the user.  Problems where a
user is only close to a single facility are not amenable to \tn and violate
\textbf{condition 3} in the definition of granularity: partitionings of the
problem where the user is not placed into the same sub-problem with the
facility closest to them would lead to a low-quality allocation. Additionally,
resource splitting cannot be used to make the problem granular, since the
objective explicitly takes into account whether facilities are used or not, and
creating multiple variables for a single <client, resource> pair would require
additional constraints \emph{across} sub-problems. More
generally, resource allocation problems where clients prefer one resource over
all other available resources by a large amount are a poor fit for \tn unless
resource splitting can be used.

\paragraph{Traffic Engineering.}

A variant of the traffic engineering problem from \S\ref{sec:prob-te} could
include hard constraints like ``flows A and B should / should not use the same
link''. This violates \textbf{condition 4} in the definition of granularity.
Randomly partitioning clients and resources into sub-problems would not work
all the time (e.g., random partitioning could drop flows A and B into different
sub-problems when flows A and B need to use the same link); smarter
partitioning algorithms can mitigate this by considering affinity between
flows, but supporting these is left to future work.

\paragraph{Global Rescheduling with Plan-Ahead.}

TetriSched~\cite{tumanov2016tetrisched} is a scheduler that can take into
account upcoming resource reservations when deciding how to allocate resources
to jobs. TetriSched allows preferences to be specified declaratively (e.g., a
job comes in at a specific start time and needs to be completed by a specific
end time). These preferences are then compiled into a mixed-integer linear
program (MILP). These MILPs can be accelerated using \tn by dividing the jobs
and resources into job and resource subsets, and solving each sub-problem
independently. However, TetriSched also supports combinatorial constraints,
such as ``a particular set of $k$ jobs must use the same resource'', which
cannot be supported by \tn without smarter partitioning algorithms.

%% file: tex/analysis.tex
\section{Analysis}

The effectiveness of POP is directly tied to how clients and resources are
partitioned across sub-problems. In this section, we consider a simple resource
allocation problem and prove that the probability of a large optimality gap
with the \tn procedure and random partitioning is low, discuss how \tn relates
to \emph{primal decomposition} (a technique used in convex optimization to
decompose certain types of optimization problems), and also note the expected
runtime benefits.

\subsection{Theoretical Analysis for a Simple Problem}
\label{sec:simple_partitioning_problem}

In settings with large numbers of clients, POP with random partitioning works
well. In this section, we consider a simplified allocation problem and compute
an upper bound on the probability that POP (using $k$ sub-problems) with random
partitioning results in a low-quality allocation.

\begin{figure}
    \center
    \includegraphics[width=0.85\columnwidth]{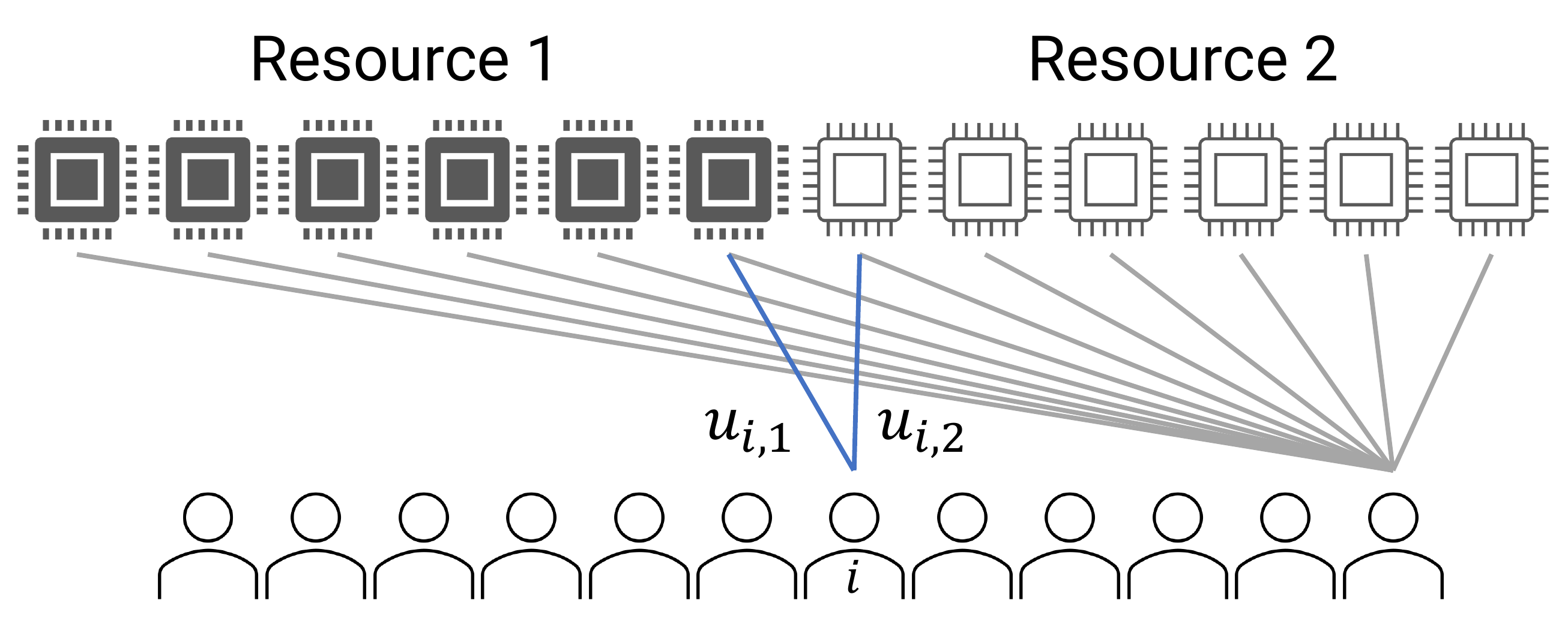}
    \caption{
        Simple partitioning problem where jobs are assigned servers (or resources).
        Each job $i$ derives utility $u_{i, 1}$ from resource 1 and
        $u_{i, 2}$ from resource 2.
        \label{fig:simple_partitioning}
    }
\end{figure}

The allocation problem we consider assigns servers to jobs. We assume that
the problem has the following properties:
\begin{itemize}

    \item $n$ jobs and servers. Each job is allocated a single server.

    \item $r$ distinct server types (equal number of each type).

    \item Job $i$ has utility $u_{i,s}$ on resource type $s$.

    \item The largest difference in utility for any job across any two servers
is $u_{\text{maxgap}}$.

\end{itemize}
A job is ``type-$s$'' if it achieves highest utility on a type-$s$ server.
With two server types,  we have type-1 and type-2 jobs (shown in
Figure~\ref{fig:simple_partitioning}).

The objective of this problem is to maximize the overall utility of the
allocation, defined as the sum of every job's utility on its assigned server.

Now, if we use \tn to solve this problem, we would equally partition servers of
each type into sub-clusters, randomly assign jobs to sub-clusters, and then
solve assignment problems separately for each sub-cluster. We wish to answer
the following questions in this regime:
\begin{enumerate}

    \item What is the optimality gap of the solution using the \tn procedure
(with respect to the optimal solution for the full problem)?

    \item How do the values of $n$, $r$, $u_{\text{maxgap}}$, and $k$ affect
this optimality gap?
\end{enumerate}

One way to quantify the optimality gap is to count the number of ``misplaced"
jobs in each sub-problem (e.g., type-1 jobs that are not assigned ``resource
1'' because there were too many other type-1 jobs in the relevant sub-problem).
Define $q_{s,t}$ to be the number of type-$s$ resources that are misplaced in
sub-problem $t$. The distance from optimal utility, i.e., optimality gap, is
bounded by the product of this number and $u_{\text{maxgap}}$ added across all
resource types and sub-problems:
\begin{align}
\text{Optimality gap} \leq \sum_{s = 1}^r \sum_{t = 1}^k q_{s,t}u_{\text{maxgap}} \label{eq-utility_bound}
\end{align}

We note that this is a loose bound for the gap, since jobs with large resource
utility gaps would be allocated their optimal resource even within a
sub-problem.

To quantify the performance gap between POP and optimal solutions, we now need
a sense of how big $q_{s,t}$ can be in practice.  We walk through the full
derivation of a bound on the probability that the optimality gap exceeds a
given value in the Appendix, but briefly sketch it here.  The random assignment
of all type-$s$ jobs to sub-problems can be interpreted as Bernoulli trials
where the probability that any given type-$r$ job is placed in a given
sub-problem is $1/k$.  We then use a classical Chernoff
bound~\cite{mitzenmacher2017probability} to compute the probability that each
$q_{s,t}$ exceeds a fraction $\delta$ of its expected value ($n/rk$).  We can
combine these across all job types and sub-problems using the union bound to
find an upper limit on the probability that the total number of misplaced jobs
exceeds $\delta n$. This allows us to bound the distance of a
randomly-partitioned POP allocation from optimal utility by $\delta
u_{\text{maxgap}} n$:
\begin{align}
\Pr\big[U(\Gamma^*) - U(\Gamma^\text{POP}) \geq \delta u_{\text{maxgap}}n \big] \leq r k \exp \Big(\frac{- \delta^2 n}{(2+\delta)rk} \Big)\label{eq-perf_bound2}
\end{align}
where $\Gamma^*$ is an optimal allocation, $\Gamma^\text{POP}$ is the
allocation returned by the \tn procedure, and $U(): \Gamma \rightarrow u$ is a
function that maps an allocation $\Gamma$ to a scalar value (the utility).

Equation~\ref{eq-perf_bound2} defines the relationship between the problem
parameters ($n$, $r$, $u_{\text{maxgap}}$ and $k$) and the probability that the
optimality gap exceeds a given fraction $\delta$ of the worst-case gap if every
job is allocated its worst resource ($u_{\text{maxgap}}n$). Concretely, the
probability decays exponentially with $n$; as the problem gets larger, the
probability of having a large optimality gap becomes very small.  The
probability also decays exponentially with $\delta^2$.  On the other hand, the
probability of a large optimality gap increases as $r$, $k$, and
$u_{\text{maxgap}}$ increase; this is to be expected, as having many
sub-problems and many resource types increases problem heterogeneity and makes
it more likely for a random partitioning to lead to misplaced jobs and a
lower-quality allocation.

To put this bound into perspective, consider a large cluster with 1 million
jobs, $k$ = 10 sub-problems, and $r$ = 4 resource types of equal amounts ($n/rk
= 25,000$); the probability that more than 3\% of jobs are not allocated their
optimal resource is upper bounded by 0.000614.

To summarize, the bound given in Equation~\ref{eq-perf_bound2} for a simple
allocation problem gives insight as to why POP works well empirically for more
complex granular resource allocation problems like those described in
\S\ref{sec:prob-instances}.

\subsection{Relationship to Primal Decomposition}
\label{sec:primal_decomposition}

For many problems, such as when the objective function is separable and convex
(that is, the objective can be expressed in the form ``$\text{Maximize } U(A) =
\sum_i U_i(A_i)$'' with per-job utility functions $U_i$), \tn can be
interpreted as the first iteration of primal decomposition, a well-known method
from convex optimization~\cite{boyd2007notes}. Primal decomposition is an
iterative technique; for a resource allocation problem, it works by decomposing
the large problem into several smaller allocation problems, each with a subset
of clients and resources. In each iteration, every sub-problem is solved
individually, and then the dual variables of each sub-problem are used to
determine how to shift resources between the sub-problems; those found to be
relatively resource-starved are given more resources from other sub-problems
for the next iteration.

Like many other techniques from the optimization literature, primal
decomposition works for a restricted set of problems, namely those with
separable objectives and certain types of constraints (see~Boyd et
al.~\cite{boyd2007notes}).  These restrictions come into effect during the
resource-shifting phase prior to subsequent iterations. For a
``well-partitioned'' problem with a separable objective (i.e., each sub-problem
has sufficient resources), one iteration of primal decomposition is often
sufficient and resource shifting is not required~\cite{boyd2007notes}. Primal
decomposition and \tn are thus equivalent for these problems, explaining why
\tn can produce a high-quality allocation efficiently. However, this
explanation does not apply to other problems where primal decomposition cannot
be used (e.g., non-convex problems, such as the MILP used in the load balancing
problem from \S\ref{sec:prob-lb}), even though we found \tn to still be
effective in such regimes.

\subsection{Expected Runtime Benefits}

We can estimate the runtime benefits of \tn when used with linear programs.
Solvers for linear programs have worst-case time complexity of $O(f(n,m)^a)$
($a \approx 2.373$~\cite{cohen2021solving} in the worst case) where $f(n,m)$ is
the number of variables ($n$ clients and $m$ resources) in the problem. If
$f(n,m)=n \cdot m$ and both clients and resources are partitioned across $k$
sub-problems, each sub-problem will have $k^2\times$ fewer variables, as
illustrated in Figure~\ref{fig:system_scaling}.  The asymptotic runtime savings
are then proportional to $k^{2a-1}$ if each sub-problem is solved serially, and
proportional to $k^{2a}$ if solved in parallel, assuming a cheap
\lstinline{reduce} step.  Some problems have an
even larger potential for runtime reduction. For example, if the allocation
considers interactions between two jobs on the same resource, then the problem
would have $n^2m$ variables, and using \tn would lead to a larger runtime
speedup (proportional to $k^{3a-1}$ if each sub-problem is solved serially,
and proportional to $k^{3a}$ if solved in parallel).

%% file: tex/implementation.tex
\section{Implementation}

\tn is easy to implement on top of a number of existing solvers for a variety
of different \apn allocation problems. The main method that needs to be
implemented is \lstinline{partition}, which given a collection of clients and
resources, assigns them to sub-problems. The subsequent \lstinline{map} step
then involves calling the existing solver routine for the already-written
problem formulation on the smaller sub-problem. The \lstinline{reduce} step is
similarly simple, and involves concatenating the allocations obtained from each
of the sub-problems and summing allocations across virtual clients and
resources (when using client and resource splitting).

We implemented \tn on top of a number of different solvers (MOSEK using
\texttt{cvxpy}~\cite{diamond2016cvxpy, agrawal2018rewriting},
Gurobi~\cite{gurobi}, and a custom solver~\cite{agrawal2021allocation} that
uses PyTorch~\cite{pytorch}) for problems across diverse domains, in $<$ 20
lines of code in each case. We implemented client splitting in about 100 lines
of Python code.

%% file: tex/evaluation.tex
\section{Evaluation}
\label{sec:evaluation}

In this section, we seek to answer the following questions: \begin{enumerate}

    \item What is the effect of \tn on allocation quality and execution time on
\apn allocation problems? How does it compare to relevant heuristics?

    \item Does \tn work across a range of solvers and types of optimization
problems?

    \item How effective are \tn's client and resource splitting optimizations
in generating high-quality allocations?

    \item How does random partitioning compare to other more sophisticated
problem partitioning strategies?

\end{enumerate}

We evaluate \tn on problems from three domains:
\begin{enumerate}

    \item \textbf{GPU cluster scheduling}, where we apply \tn to solve the
optimization problems used in Gavel (\S\ref{sec:prob-cs}), and compare with the
greedy Gandiva policy~\cite{xiao2018gandiva}.

    \item \textbf{Traffic engineering} across Wide Area Networks, where we
apply \tn to solve the problem formulations in \S\ref{sec:prob-te}, and compare
to CSPF and NCFlow \cite{abuzaid2021contracting}.

    \item \textbf{Shard load balancing} in distributed storage systems, where
we apply \tn on the problem formulation in \S\ref{sec:prob-lb}, and compare to
a heuristic from E-Store~\cite{taft2014store}.

\end{enumerate}

Where relevant, we integrate \tn into systems such as
Gavel~\cite{narayanan2020heterogeneity} to measure the end-to-end impact of \tn
on application performance.  Our results span three different cluster
scheduling policies (max-min fairness, minimize makespan, and proportional
fairness), two traffic engineering policies (maximize total flow, and maximize
concurrent flow), and one load balancing policy (minimize number of shard
transfers as load changes).

We first present end-to-end experiments, then present some microbenchmarks that
examine the impact of various algorithmic contributions in \tn.

\subsection{End-to-End Results}

We first demonstrate \tn's end-to-end effectiveness on various problems.  We
compare to approaches based on allocation quality, and time needed to compute
the allocation; the runtime for \tn includes the runtime for solving the
optimization problems for sub-problems.  In all of our experiments, ``Exact
sol.'' is the original unpartitioned problem formulation and solver used by the
reference system (e.g., Gavel for cluster scheduling). We believe this is a
fair baseline since it represents what people use today if using optimization
problem formulations for resource allocation. We use the same evaluation
methodology as related work.  The total number of threads given to solvers for
our baselines and \tn are the \emph{same}. If $k$ sub-problems are solved in
parallel when using \tn, each sub-problem uses $1/k$ of the number of threads.
We also present heuristics where relevant. Unfortunately, not every problem has
a state-of-the-art heuristic. For example, it is not clear how to use a
heuristic to solve for an approximate proportionally-fair allocation.  We
explicitly note when we use client or resource splitting.

\begin{figure}
    \center
    \includegraphics[width=0.9\columnwidth]{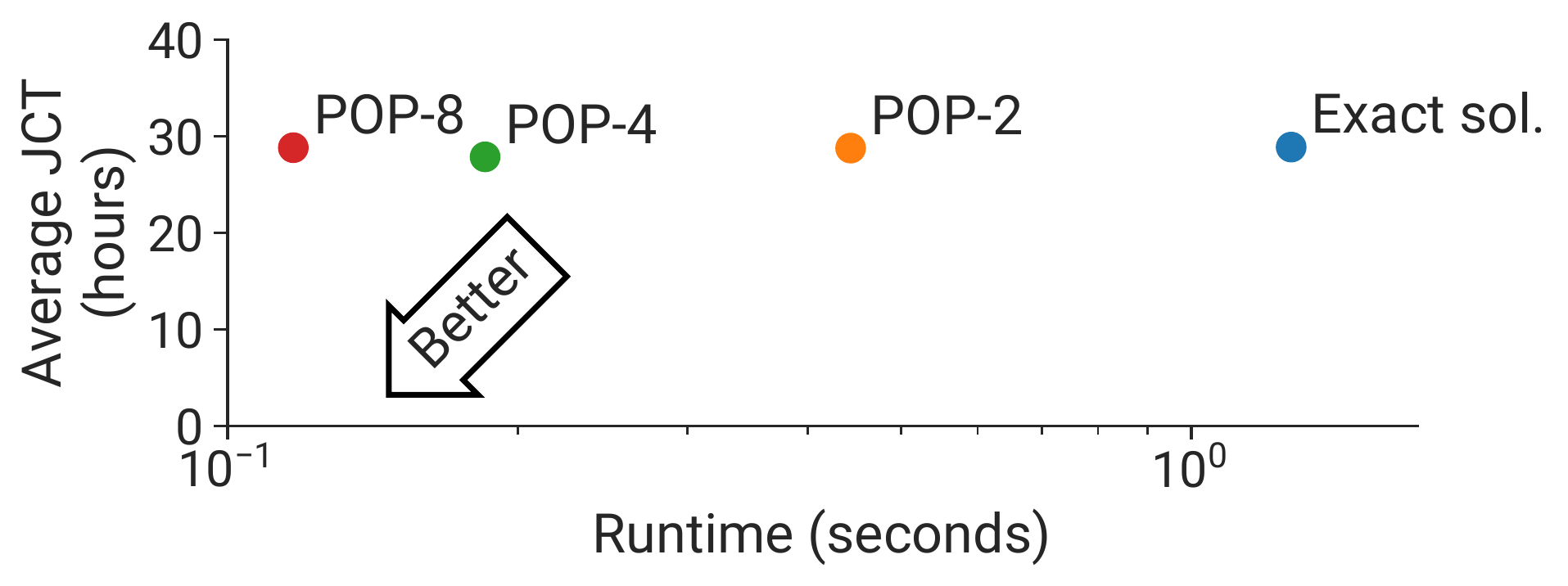}
    \vspace{-0.1in}
    \caption{
        Results for the max-min fairness policy (with space sharing) for
        cluster scheduling for the formulation shown in \S\ref{sec:prob-cs}
        (``Exact sol.'') and its \tn variants. \tn-$k$ uses $k$ sub-problems.
        \label{fig:max_min_fairness_packed_full_run}
    }
\end{figure}

\subsubsection{Cluster Scheduling}

We used \tn to accelerate various cluster scheduling policies supported by
Gavel~\cite{narayanan2020heterogeneity}. We then used these \tn-ped policies in
Gavel's full simulator\footnote{The Gavel
paper~\cite{narayanan2020heterogeneity} shows that its simulator demonstrates
performance very similar to behavior on the physical cluster.} to measure the
impact of \tn on end-to-end metrics of interest, like average job completion
time and makespan for real traces.  The traces and methodology used are
identical to those used in Gavel.

\paragraph{Max-Min Fairness.} We show the trade-off between runtime and
allocation quality for the max-min fairness policy with space sharing on a
large problem (2048$^2$ job pairs on a 1536-GPU cluster) in
Figure~\ref{fig:max_min_fairness_effective_throughput_ratios_and_runtime} (in
the introduction). \tn leads to an extremely small change in the average
effective throughputs across all jobs ($<$ 1\%), with a 22.7$\times$
improvement in runtime.  Gandiva~\cite{xiao2018gandiva}, on the other hand,
uses a heuristic to assign resources to job pairs, resulting in $1.9\times$
worse allocation quality.

We unfortunately could not run end-to-end simulations for such large problem
sizes: the simulation involves running thousands of allocation problems, since
an allocation problem needs to be solved every time a new job arrives at the
cluster or an old job completes. This would take months to run at scale by
virtue of the number of problems that need to be solved and the time taken for
each problem. Instead, we show full simulation results on more moderate problem
sizes.  These experiments involve dynamic changes: the full simulation involves
new jobs coming in and old jobs completing, and consequently the set of jobs is
not static.

We ran experiments with 96 GPUs (32 V100, P100, and K80 GPUs). The original
heterogeneity-aware Least Attained Service policy without space sharing has a
small number of variables (on the order of hundreds). Even on such smaller
problem sizes, the quality of allocation with \tn is high, with only up to a
$5\%$ drop in average JCT (not pictured).

Figure~\ref{fig:max_min_fairness_packed_full_run} shows the average JCT of the
original Least Attained Service policy from \S\ref{sec:prob-cs}, with space
sharing, along with three \tn-ified versions using 2, 4, and 8 sub-problems.
With space sharing, the number of variables scales quadratically with the
number of jobs: this leads to a performance speedup of 11$\times$ with $k=8$
compared to the full problem formulation, and similar average JCT.

We see similar behavior for max-min fairness policies when clients have more
attributes (e.g., different priority levels).  Average JCTs are almost
identical when jobs request multiple GPUs, and increase by 5\% for
high-priority jobs in workloads containing a mix of low- and high-priority
jobs, using the Gavel simulator as before.

\begin{figure}
    \center
    \includegraphics[width=0.85\columnwidth]{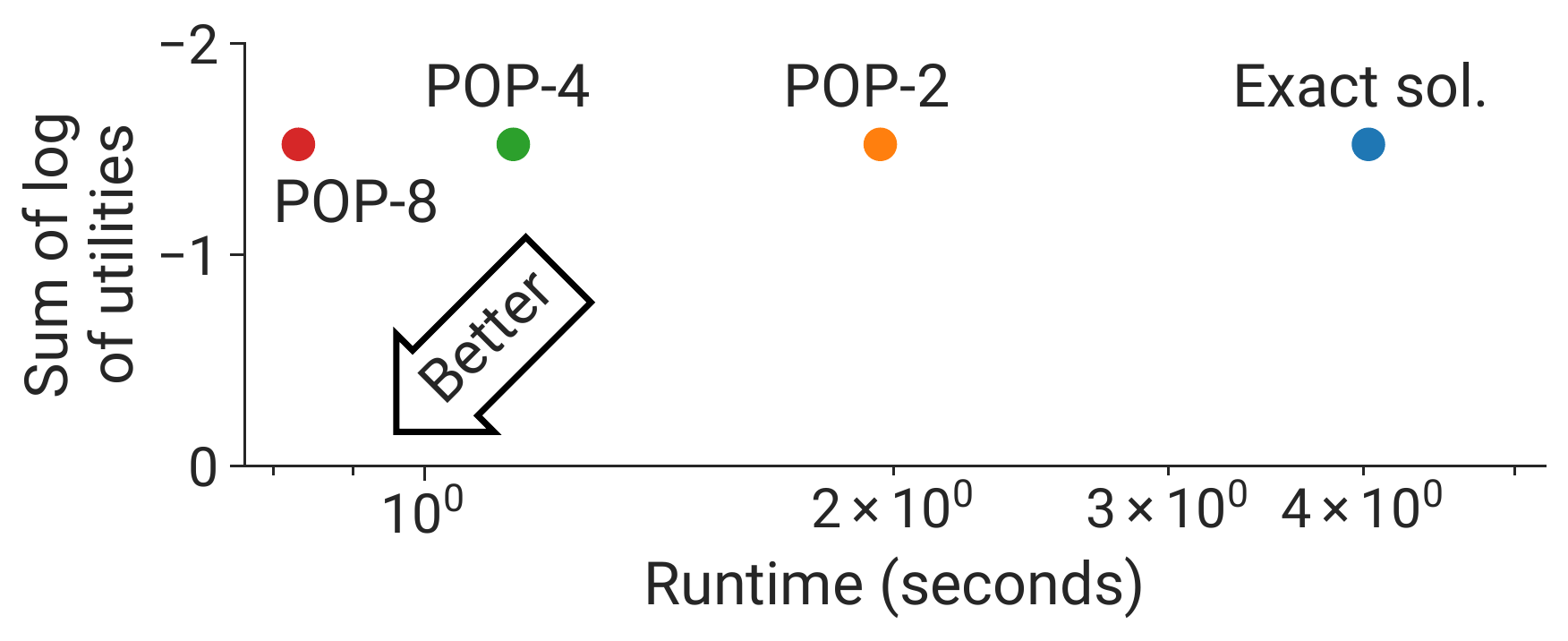}
    \vspace{-0.1in}
    \caption{
        Results for the proportional fairness policy for cluster scheduling for
        the formulation shown in \S\ref{sec:prob-cs} (``Exact sol.'') and its
        \tn variants. \tn-$k$ uses $k$ sub-problems.
        \label{fig:proportional_fairness}
    }
\end{figure}

\paragraph{Proportional Fairness.} We ran a simple experiment with the
proportional fairness policy with $10^6$ jobs and a similar number of
resources.  Figure~\ref{fig:proportional_fairness} shows \tn combined with a
proportional fairness policy. This allocation problem is a general convex
optimization problem (not a linear program), with a sum-of-$\log$ objective.
For this problem, we implement \tn on top of a custom
solver~\cite{agrawal2021allocation} that runs an order of magnitude faster than
commercial solvers for this particular problem formulation.  We see strong
scaling performance as we increase the number of sub-problems (4.9$\times$
reduction in runtime with 8 sub-problems), with an extremely small optimality
gap ($7\times10^{-5}$). 

\begin{figure}
    \center
    \includegraphics[width=0.9\columnwidth]{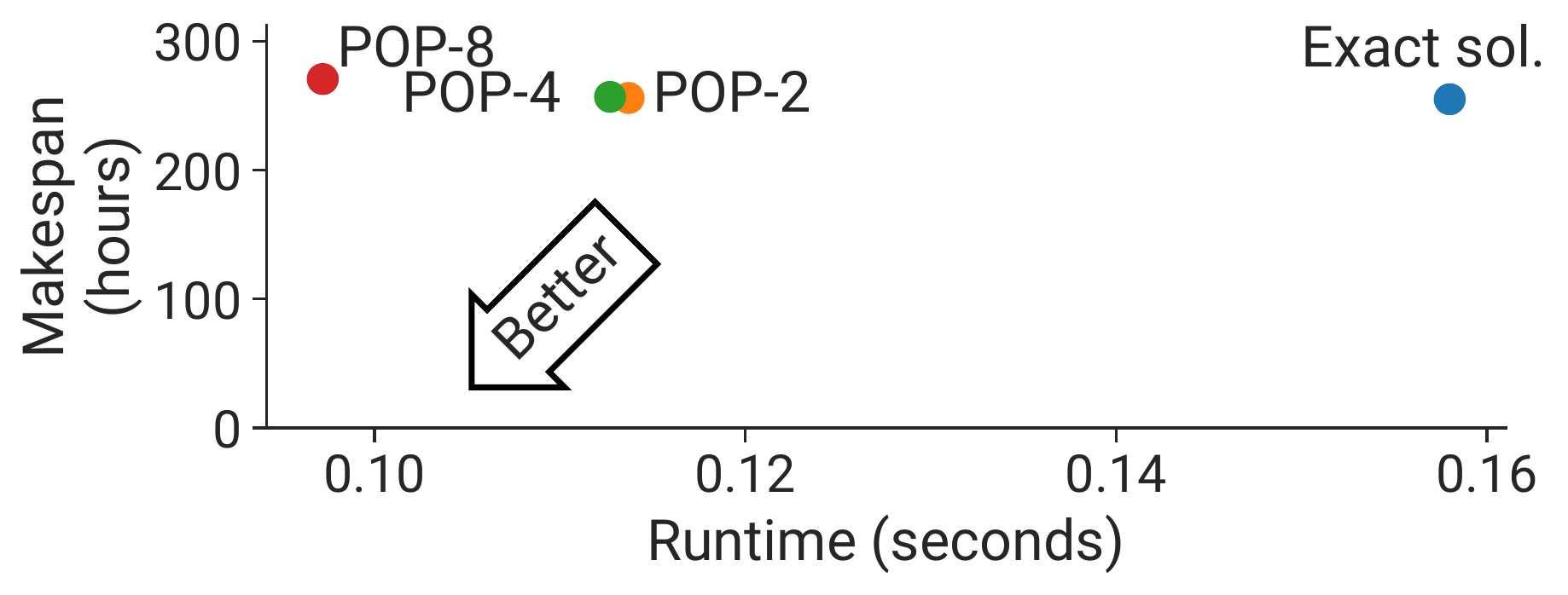}
    \vspace{-0.1in}
    \caption{
        Results for the minimize makespan policy for cluster scheduling for the
        formulation shown in \S\ref{sec:prob-cs} (``Exact sol.'') and its \tn
        variants. \tn-$k$ uses $k$ sub-problems.
        \label{fig:minimize_makespan_full_run}
    }
\end{figure}

\paragraph{Minimize Makespan.} Figure~\ref{fig:minimize_makespan_full_run}
shows the makespan of variants of the ``minimize makespan'' policy. This policy
again is a simple linear program with number of variables linear in the number
of jobs and resource types.  Consequently, the runtime improvements are lower
($1.6\times$), but the end-to-end makespan over the trace is nearly identical.

\begin{table}[t!]
\centering
{\footnotesize
  \begin{tabularx}{0.8\columnwidth}{@{\hspace{12pt}}l||@{\hspace{12pt}}XX@{}}
\toprule
  {\bf Topology} & {\bf \# Nodes} & {\bf \# Edges} \\ \midrule
  \href{http://www.topology-zoo.org/maps/Kdl.jpg}{Kdl}                     & $754$       & $1790$ \\
  \href{http://www.topology-zoo.org/maps/Cogentco.jpg}{Cogentco}           & $197$       & $486$  \\
  \href{http://www.topology-zoo.org/maps/UsCarrier.jpg}{UsCarrier}         & $158$       & $378$  \\
  \href{http://www.topology-zoo.org/maps/Colt.jpg}{Colt}                   & $153$       & $354$  \\
  \href{http://www.topology-zoo.org/maps/GtsCe.jpg}{GtsCe}                 & $149$       & $386$  \\
  \href{http://www.topology-zoo.org/maps/TataNld.jpg}{TataNld}             & $145$       & $372$  \\
  \href{http://www.topology-zoo.org/maps/DialtelecomCz.jpg}{DialtelecomCz} & $138$       & $302$  \\
  \href{http://www.topology-zoo.org/maps/Deltacom.jpg}{Deltacom}           & $113$       & $322$  \\
\end{tabularx}
}
\caption{The WAN topologies used to benchmark \tn for traffic engineering problems, obtained from Internet Topology Zoo~\cite{knight2011internet}.}
\vspace{-0.2in}
\label{t:topologies}
\end{table}

\begin{figure}
    \center
    \includegraphics[width=0.88\columnwidth]{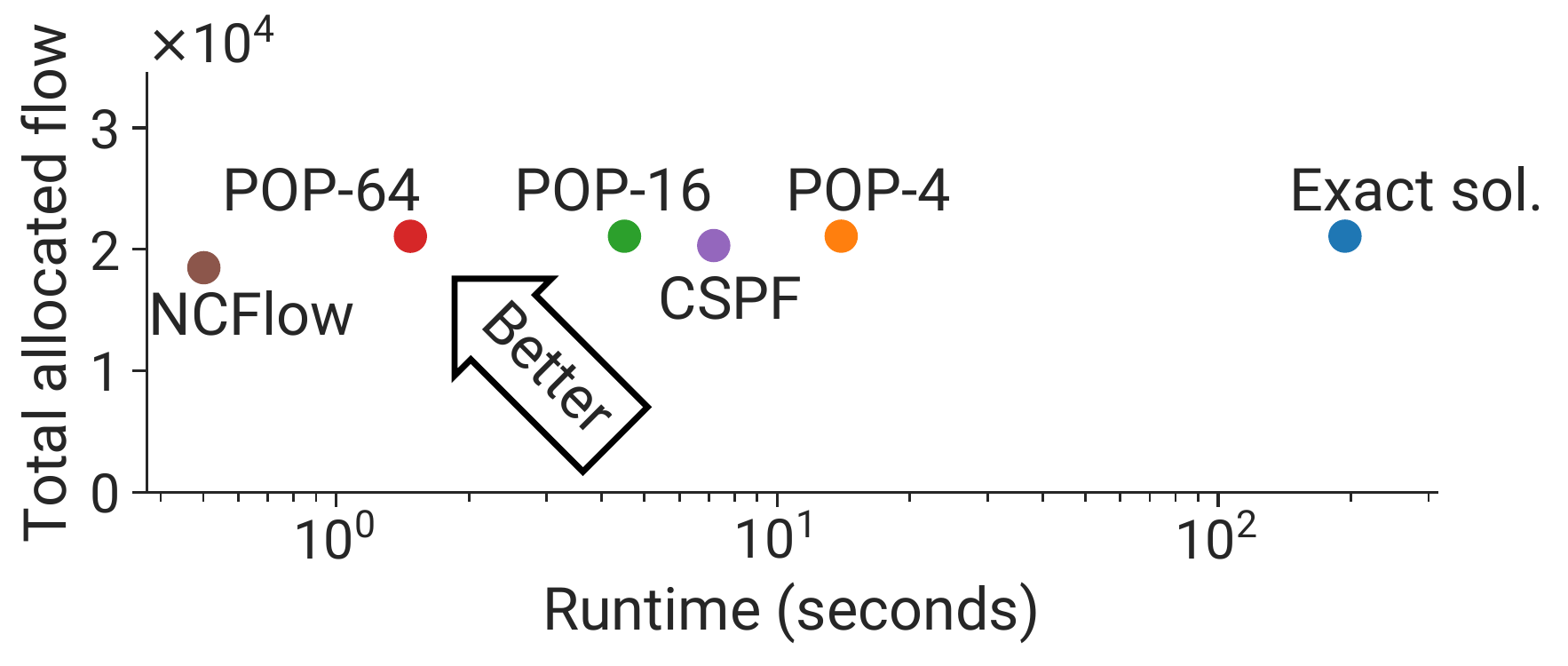}
    \vspace{-0.1in}
    \caption{
        Results for the max-flow problem for traffic engineering for a single
        topology and traffic matrix. The scatterplot shows runtimes and total
        allocated flow for the formulation shown in \S\ref{sec:prob-te}
        (``Exact sol.'') and its \tn variants, as well as CSPF and NCFlow.
        \label{fig:total_flow_and_runtime}
        \vspace{-0.1in}
    }
\end{figure}

\subsubsection{Traffic Engineering}

We tested \tn on several large networks (shown in Table~\ref{t:topologies})
from the Topology Zoo repository \cite{knight2011internet}, with similar
results. For each topology, we benchmarked \tn on sets of synthetic traffic
matrices, which were generated using several traffic models: {\sf
Gravity}~\cite{Cohen2003,gravity}, {\sf Uniform}, {\sf
Bimodal}~\cite{Cohen2003}, and {\sf Poisson}. These traffic matrices were
previously used in NCFlow\footnote{The full set of traffic matrices can be
found here:
\url{https://github.com/netcontract/ncflow}.}~\cite{abuzaid2021contracting}.
{\sf Poisson} represents a skewed workload, where a small percentage of
commodities dominate the network demand. For this workload, we use the
client-splitting algorithm from \S\ref{sec:client_and_resource_splitting} to
improve allocation quality. We do not use the client-splitting algorithm for
the other traffic matrices.

\paragraph{Total Flow.} Figure~\ref{fig:total_flow_and_runtime} shows the
trade-off between runtime and allocated flow on the Kentucky Data Link network
({\sf Kdl} in Table~\ref{t:topologies}), which has 754 nodes and 1790 edges
spanning the Eastern half of continental USA. We instantiated over
$5\times10^5$ demands to up to 4 paths in the network.  The flow allocated by
\tn is within 1.5\% of optimal when using 64 sub-problems, yet $100\times$
faster than the original problem. We also compare favourably to the Constrained
Shortest Path First (CSPF) heuristic~\cite{fortz2002traffic} and the
recently-published NCFlow~\cite{abuzaid2021contracting}.  Note that NCFlow is
\emph{not} a heuristic, but a state-of-the-art approach that uses a problem
decomposition technique explicitly tuned for the max-flow problem.

\begin{figure}
    \center
    \includegraphics[width=0.88\columnwidth]{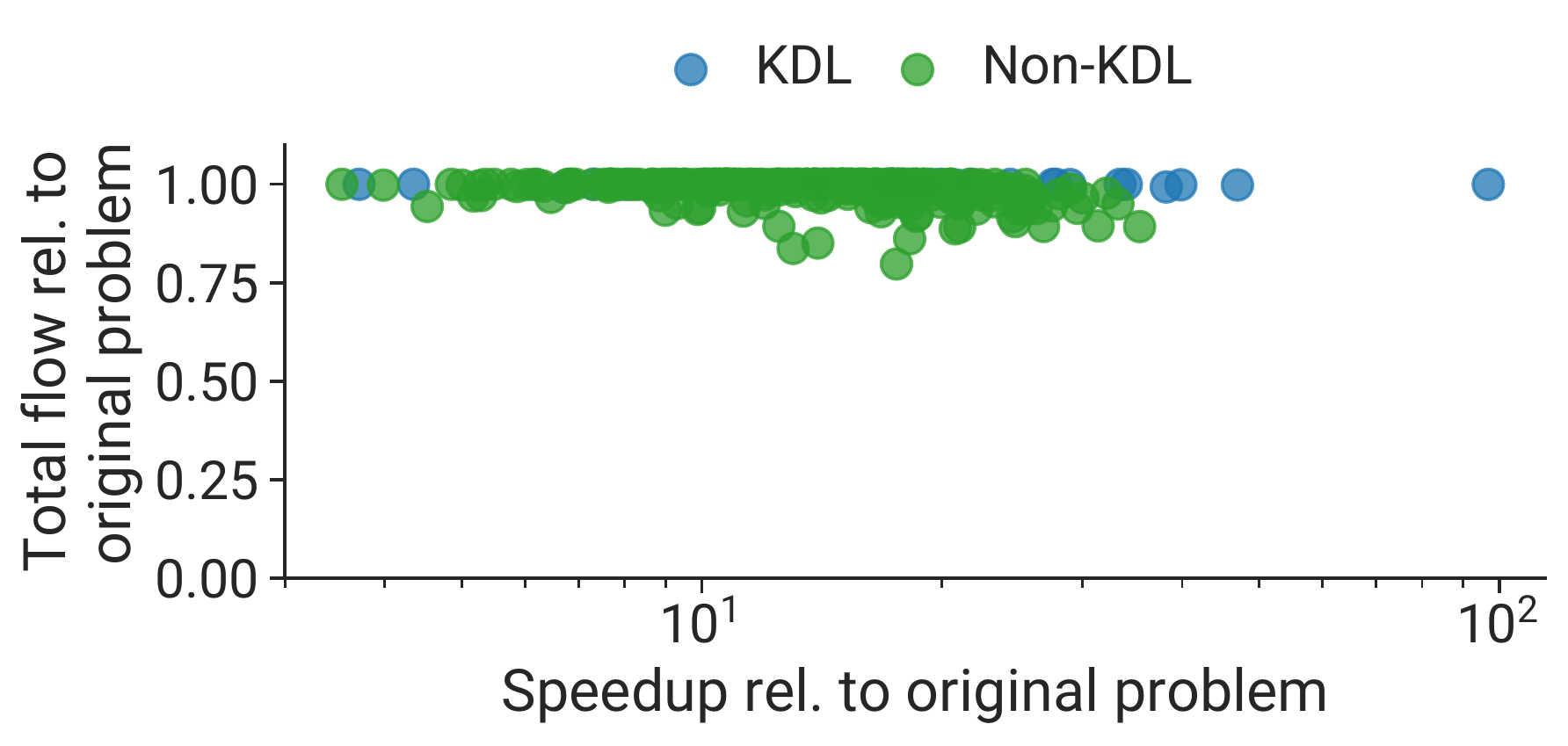}
    \vspace{-0.1in}
    \caption{
        Results for the max-flow problem for traffic engineering for multiple
        topologies and traffic matrices. The scatterplot shows runtimes and
        allocated total flow for \tn-16 across 275 experiments, separated by
        large ({\sf kdl}) and small (non-{\sf kdl}) topologies.
        \label{fig:total_flow_scatterplot}
    }
\end{figure}

Figure~\ref{fig:total_flow_scatterplot} shows the improvement in allocation
quality and runtime compared to the original LP formulation presented in
\S\ref{sec:prob-instances} with \tn using 16 sub-problems. Each point in the
scatterplot represents a different topology and traffic matrix. We see larger
speedups for the larger {\sf Kdl} topology. We used client splitting with a
threshold ($t$) of 0.75 for the {\sf Poisson} traffic matrices (where some
commodities have large demands), and no client splitting for the other traffic
models, which were granular out of the box. As stated before, resource
splitting was used for all traffic matrices to ensure that each
sub-problem has all links (but with lower capacity).

\begin{figure}
    \center
    \includegraphics[width=0.95\columnwidth]{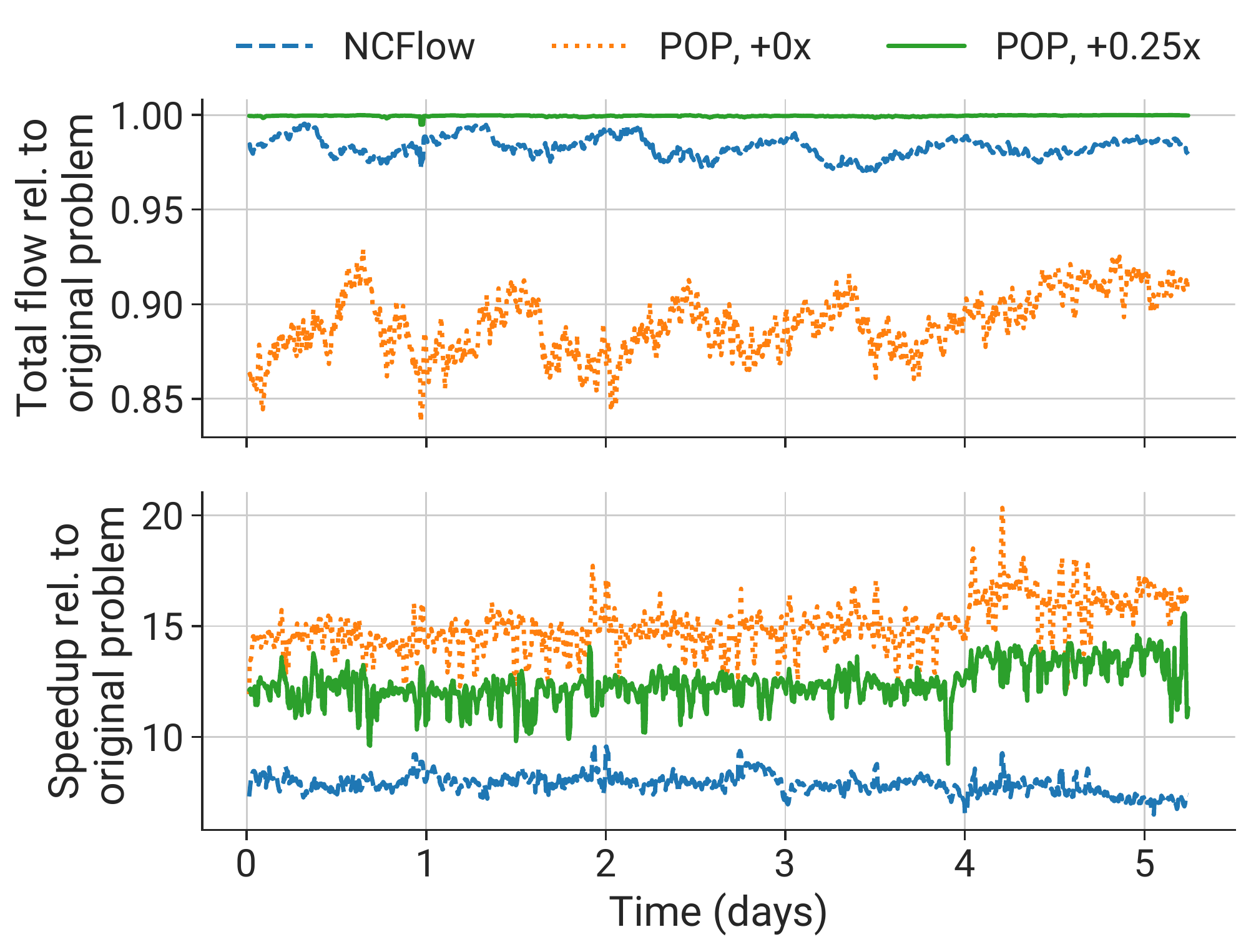}
    \vspace{-0.1in}
    \caption[\tn on Traffic Measured in a Production WAN]{Allocated flow
    and speedup relative to original problem on a 5-day sequence of real-world
    traffic matrices from a private WAN with 100s of nodes and edges. With
    client splitting ($t=0.25$), \tn allocates $>$99\% of the total flow with
    a $12.5\times$ median speedup.
    }
    \label{fig:private-wan}
\end{figure}

We also ran experiments with a sequence of \emph{real-world} traffic traces
collected on a private industrial WAN with hundreds of nodes and edges.
\autoref{fig:private-wan} plots the moving average (over 5 windows) of the
total flow and speedup relative to the original problem for NCFlow, \tn with no
client splitting, and \tn with $t=0.25$ client splitting.  Without client
splitting, \tn achieves significant speedups ($15\times$ in the median case)
compared to the original problem, but allocates $89.1$\% of the total flow in
the median case. However, \tn with client splitting nearly matches the total
flow allocated in the original problem ($99.9$\% in the median case), while
still achieving a median $12.5\times$ speedup.

\begin{figure}
    \center
    \includegraphics[width=0.93\columnwidth]{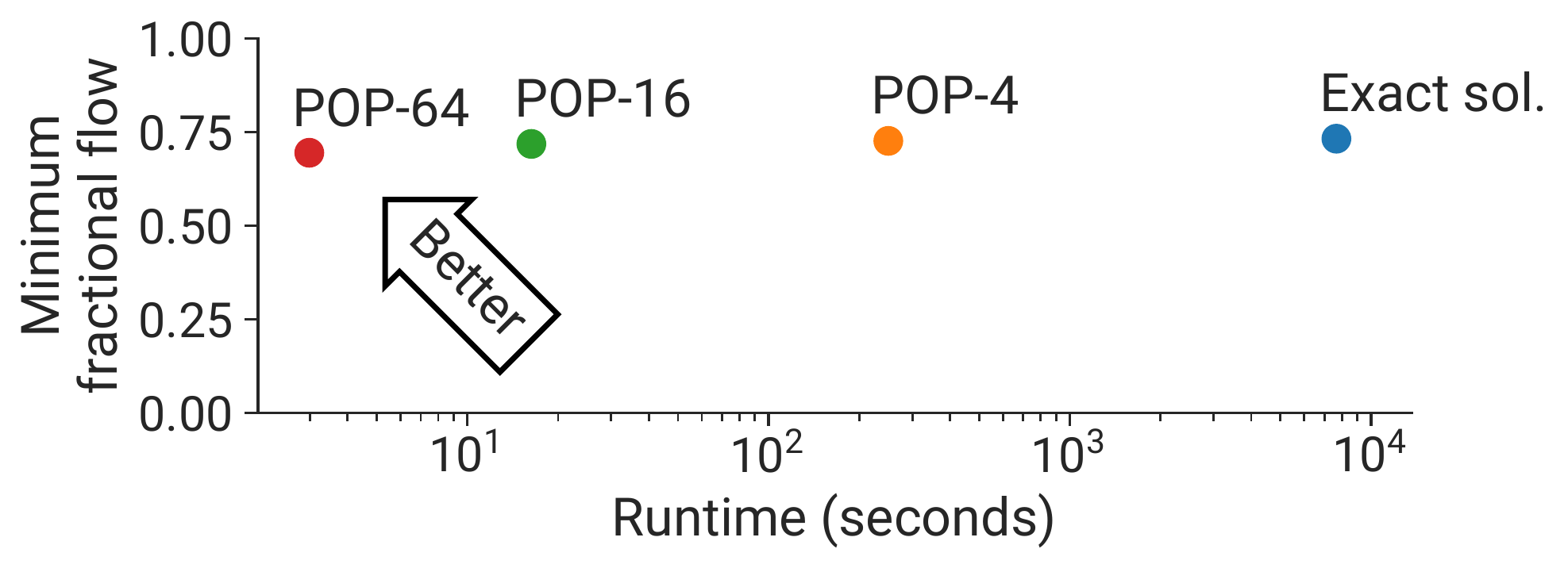}
    \vspace{-0.1in}
    \caption{
        Results for the maximum concurrent flow problem for traffic engineering
        for a single topology and traffic matrix. The scatterplot shows
        runtimes and minimum fractional flow for the formulation shown
        in \S\ref{sec:prob-te} (``Exact sol.'') and its \tn variants.
        \label{fig:min_fractional_flow_and_runtime}
    }
\end{figure}

\paragraph{Maximum Concurrent Flow.} Similarly, we benchmarked \tn on the
maximum concurrent flow objective using the same set of topologies and traffic
matrices.  Figure~\ref{fig:min_fractional_flow_and_runtime} shows the trade-off
between runtime and minimum fractional flow on the {\sf Kdl} topology, using
the same traffic matrix in Figure~\ref{fig:total_flow_and_runtime}. The
objective value realized by \tn is again within 1.5\% of optimal when using 64
sub-problems, yet $1000\times$ faster than the original problem. As before, we
use client splitting with a threshold of 75\% for the {\sf Poisson} traffic
matrices, and no client splitting for the other traffic matrices. Resource
splitting is used for all traffic matrices.

\begin{figure}
    \center
    \includegraphics[width=0.9\columnwidth]{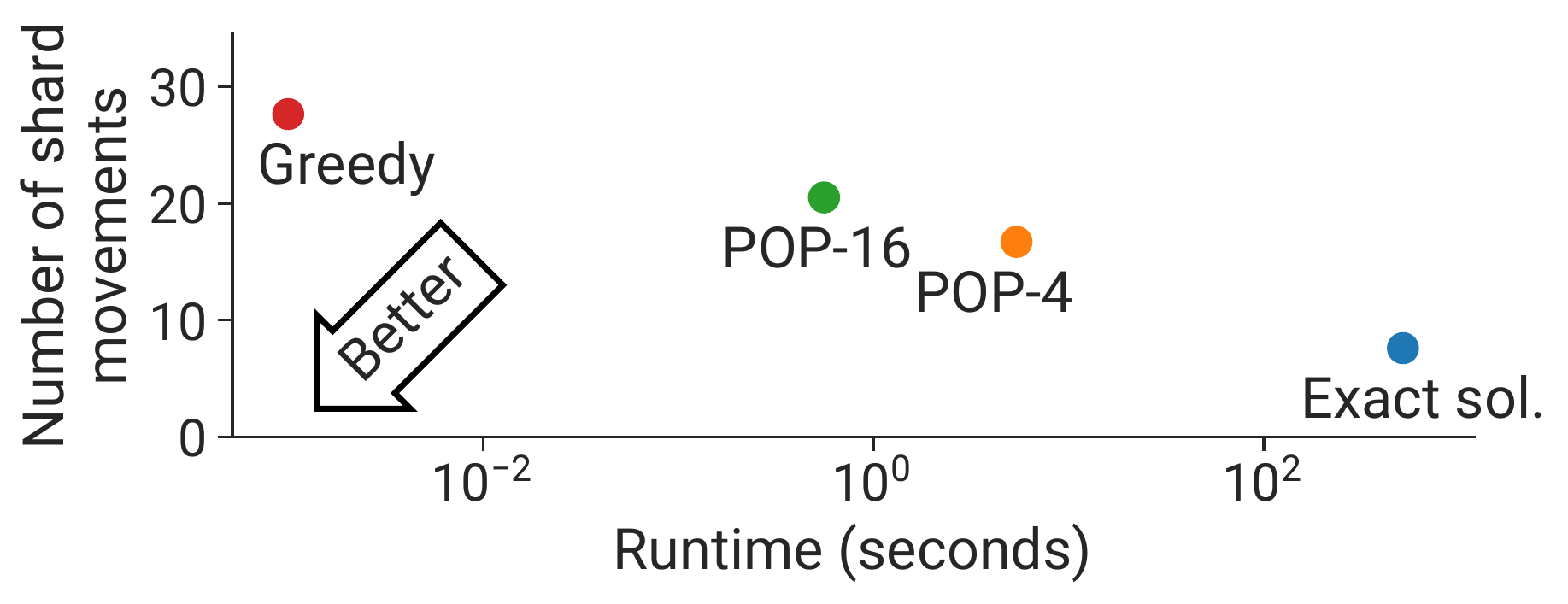}
    \vspace{-0.1in}
    \caption{
        Results for the minimize shard movement policy for load balancing for
        the formulation shown in \S\ref{sec:prob-lb} (``Exact sol.'') and its
        \tn variants. We also compare to a greedy heuristic (``Greedy'').
        \tn-$k$ uses $k$ sub-problems.
        \label{fig:number_of_shard_movements_and_runtime}
    }
    \vspace{-0.1in}
\end{figure}

\subsubsection{Load Balancing}

In Figure~\ref{fig:number_of_shard_movements_and_runtime}, we evaluate \tn on a
load balancing problem.  In the problem, we have 1024 shards of data each
assigned to exactly one of 64 servers.  Each round, we receive the query load
of each shard and compute a new assignment of shards to servers such that each
server has approximately (within 5\%) the same amount of load across its shards
but the number of shard movements is minimized.  We examine the performance of
\tn with various numbers of sub-problems and compare it to the original
optimization problem (\S\ref{sec:prob-lb}) and a greedy heuristic algorithm
from E-Store~\cite{taft2014store}.  For each system, we run 100 rounds of the
problem. In each round, we generating a new load distribution and rerun the
load balancing algorithm. We report the average runtime and number of shard
movements across these rounds.

We find that \tn improves the runtime over the original problem by two orders
of magnitude, while outperforming the greedy heuristic. The exponential scaling
of MILP solvers restricted us to smaller problem sizes for the purpose of
comparing against the optimal solution. Since shard movements are stateful
(previous round's solution is initial state for current round), we added an
extra step to re-balance the aggregate load in the relatively small
sub-problems, requiring a few extra shard movements. As $k$ increases, the
number of sub-problems and thus the number of these movements also increases,
which is why \tn-$k$ does worse as $k$ increases. This becomes less of an issue
for larger problem sizes where random allocations are likely to remain
balanced.

\subsection{Effectiveness of Client and Resource Splitting}

\begin{figure}
    \center
    \begin{subfigure}{\columnwidth}
        \center
        \includegraphics[width=\columnwidth]{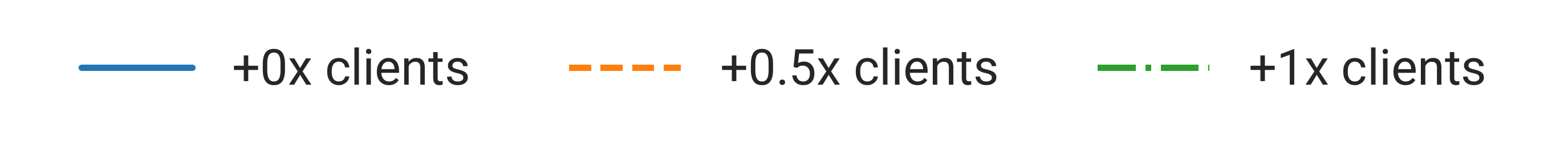}
        \vspace{-0.2in}
    \end{subfigure}
    \begin{subfigure}[l]{0.49\columnwidth}
        \center
        \includegraphics[width=\columnwidth]{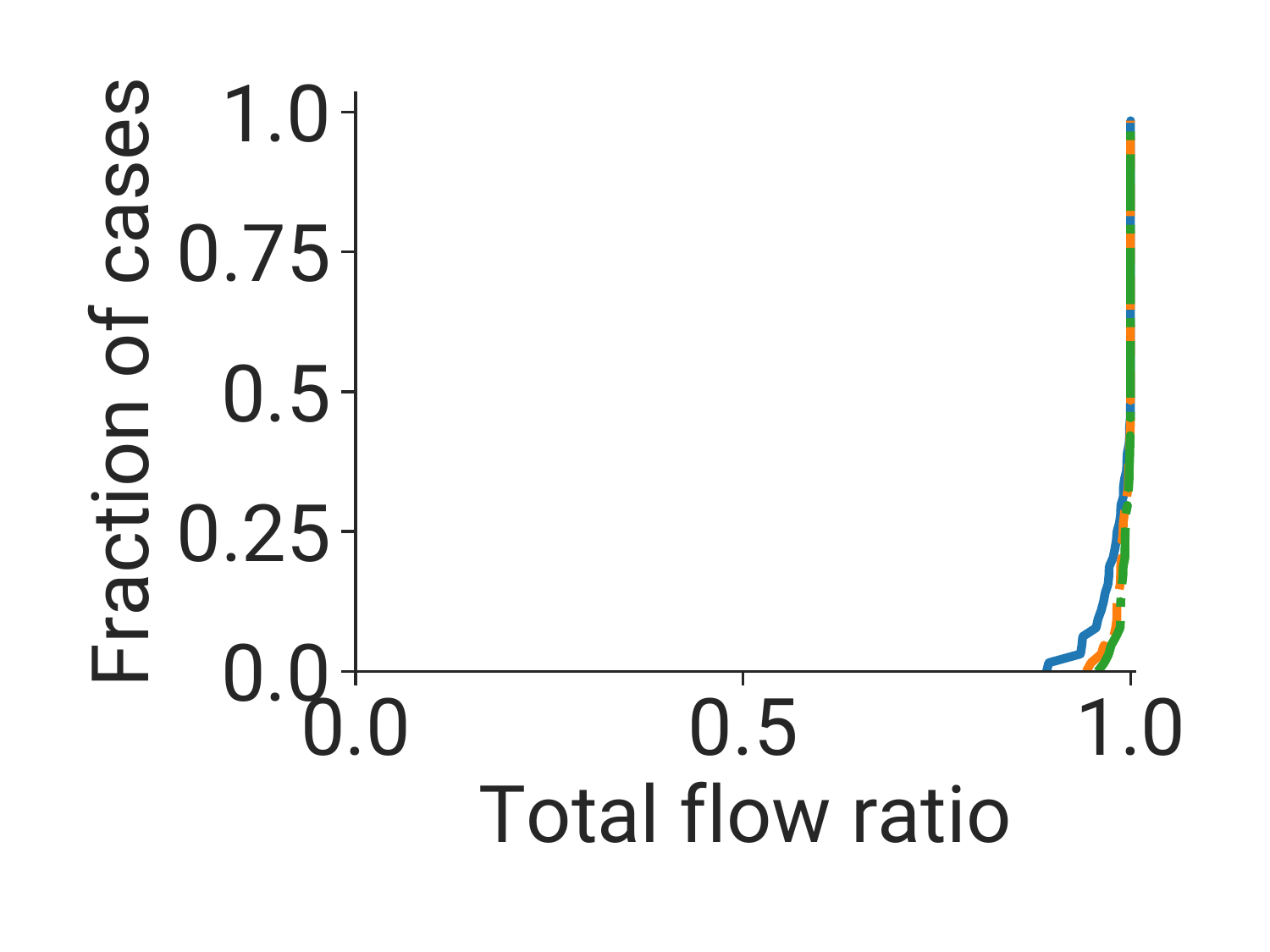}
        \vspace{-0.2in}
        \caption{Gravity (total flow ratio).}
    \end{subfigure}
    \begin{subfigure}[l]{0.49\columnwidth}
        \center
        \includegraphics[width=\columnwidth]{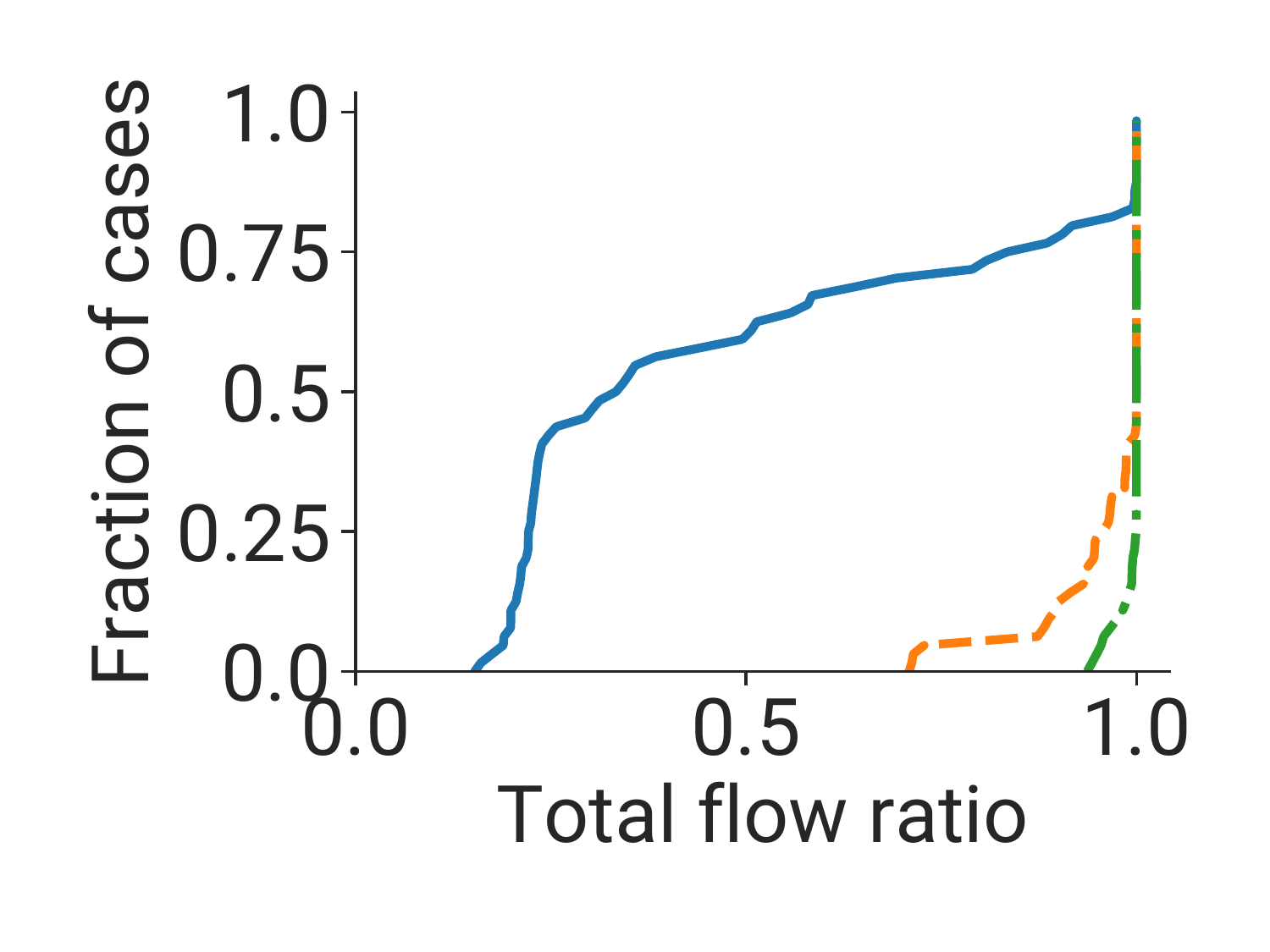}
        \vspace{-0.2in}
        \caption{Poisson (total flow ratio).}
    \end{subfigure}
    \begin{subfigure}[l]{0.49\columnwidth}
        \center
        \includegraphics[width=\columnwidth]{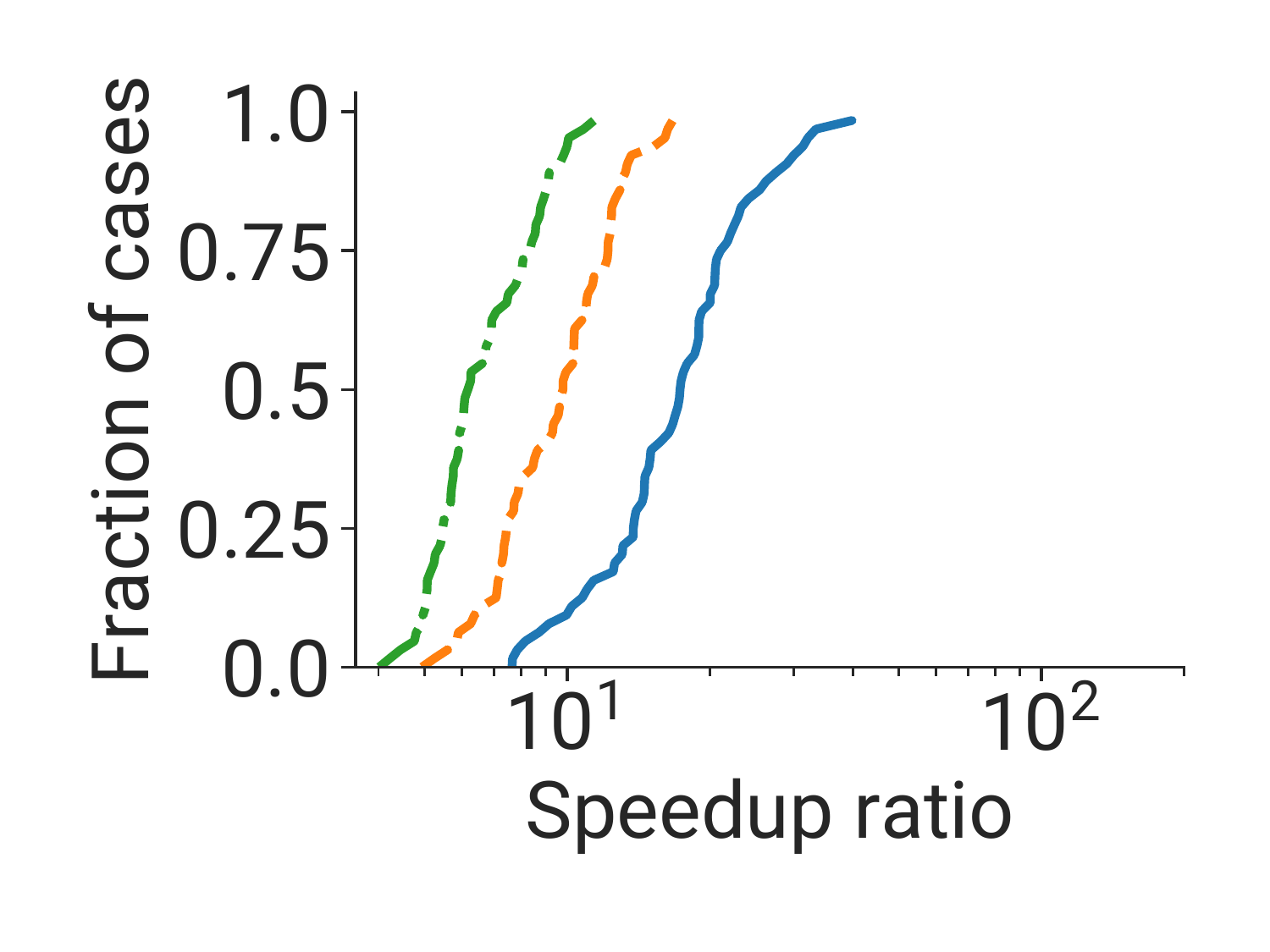}
        \vspace{-0.2in}
        \caption{Gravity (speedup ratio).}
    \end{subfigure}
    \begin{subfigure}[l]{0.49\columnwidth}
        \center
        \includegraphics[width=\columnwidth]{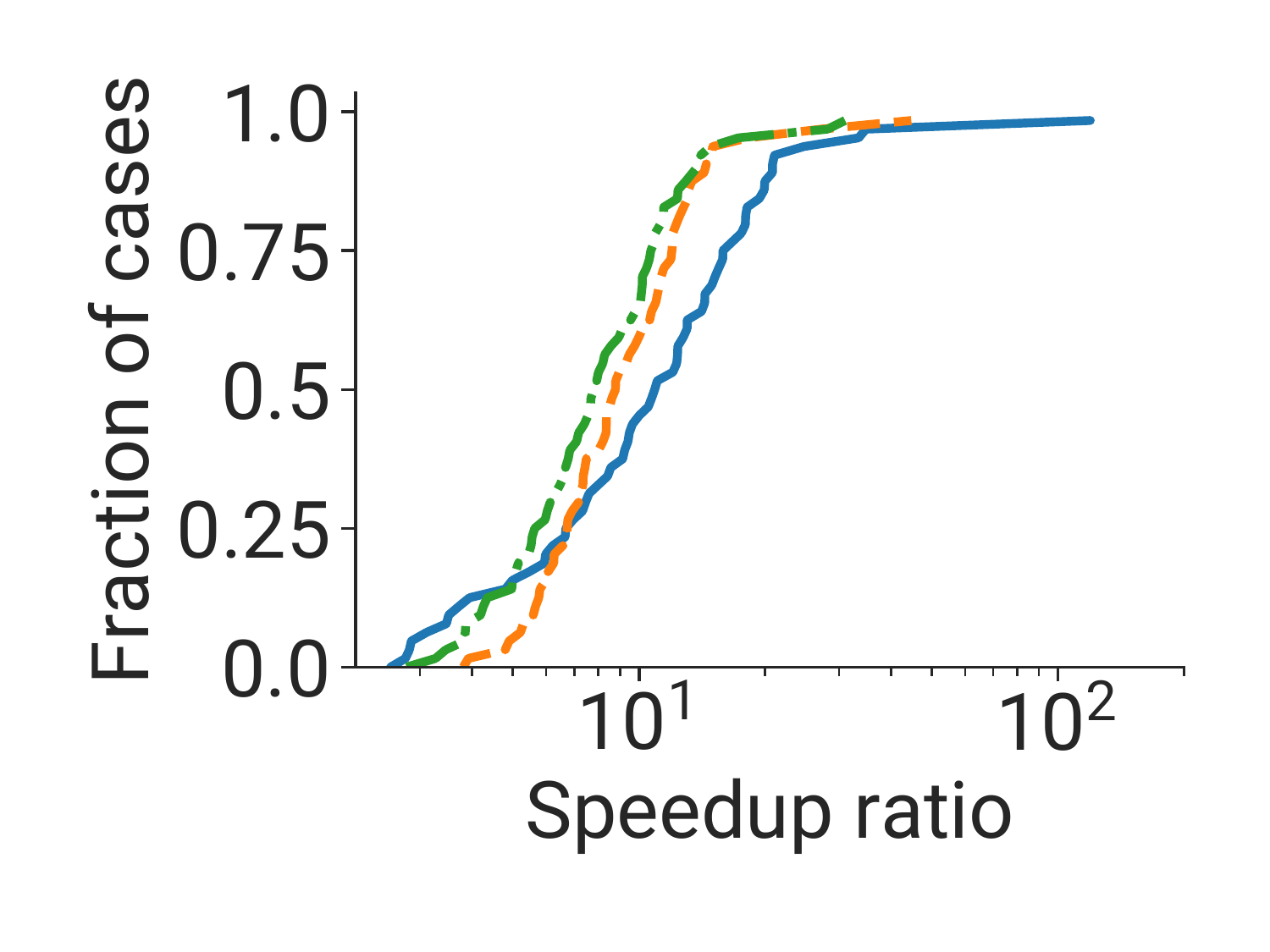}
        \vspace{-0.2in}
        \caption{Poisson (speedup ratio).}
    \end{subfigure}
    \caption{
        Comparison of \tn-16 relative to original problem for the max-flow
        objective in traffic engineering, across different levels of additional
        split clients ($0\times$ to $1\times$) and traffic matrices from two
        traffic models: {\sf Gravity} and {\sf Poisson} (which is skewed).
        \label{fig:TE_client_split_sweep}
    }
\end{figure}

\begin{figure}
    \center
    \includegraphics[width=0.8\columnwidth]{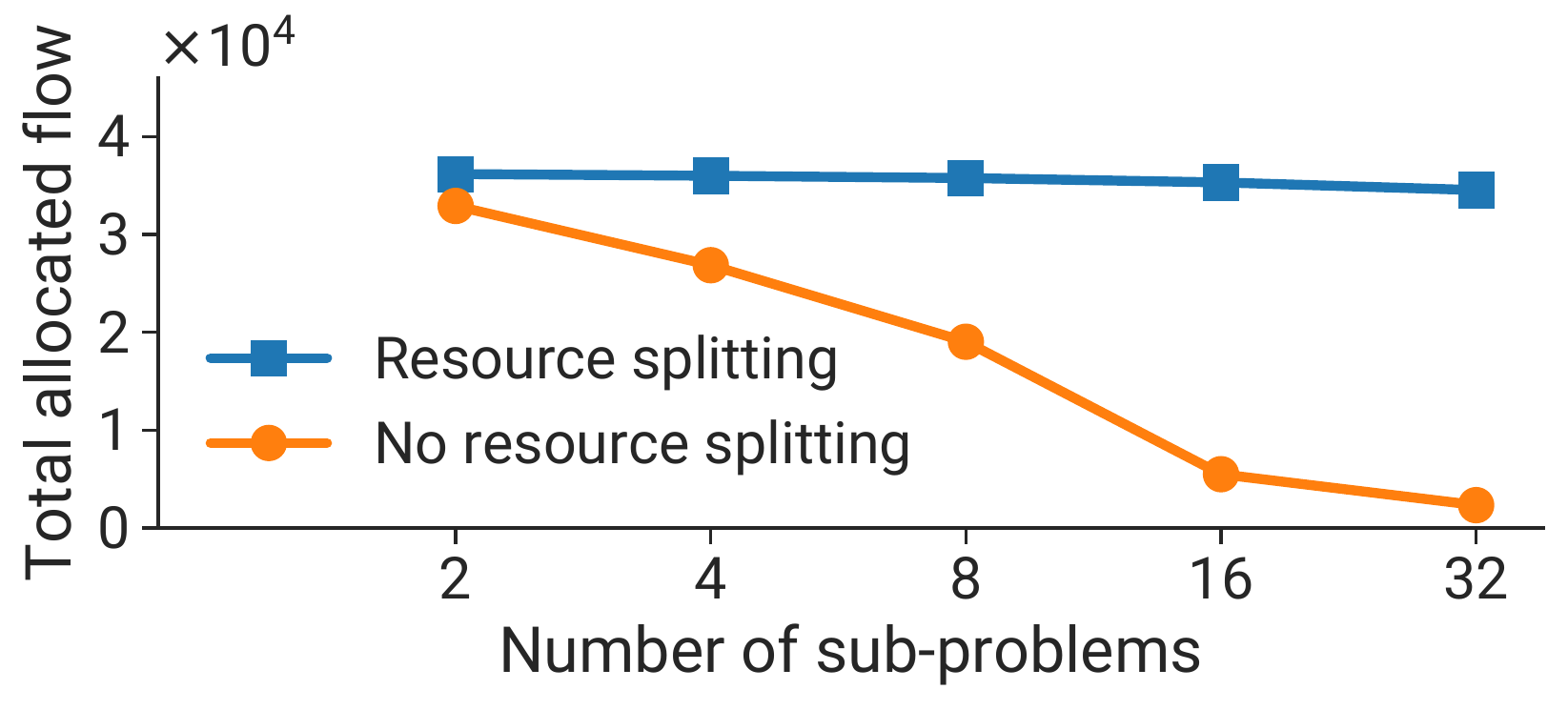}
    \vspace{-0.1in}
    \caption{
        Effect of resource splitting on total flow as the number of
        sub-problems ($k$) is adjusted for a TE problem with a max-flow
        objective (Cogentco topology and {\sf Gravity} traffic matrix).
    }
    \label{fig:TE_resource_split_sweep}
    \vspace{-0.1in}
\end{figure}

Figure~\ref{fig:TE_client_split_sweep} shows the effect of client splitting on
total flow and runtime when using \tn with 16 sub-problems, on a traffic
engineering problem with ``large'' clients ({\sf Poisson} traffic model) as
well as a more typical set of clients ({\sf Gravity} traffic model) and a
max-flow objective. The figure shows separate cumulative distributions of
approximately 100 different experiments for each traffic model and client
splitting threshold ($t$ in Algorithm~\ref{alg:pop_procedure}).

We see that with skewed traffic ({\sf Poisson} traffic model) and no client
splitting, the total flow is typically far from optimal. Client splitting
drastically increases the median relative total flow from 0.2 to near 1.0 for
these problems, at the cost of some runtime overhead (due to an increase in the
number of variables).  In contrast, the problems with {\sf Gravity} traffic get
near-optimal allocated flow without client splitting.

Figure~\ref{fig:TE_resource_split_sweep} shows the effect of resource splitting
on total flow when using \tn with various numbers of sub-problems ($k$) on the
Cogentco topology and {\sf Gravity} traffic model.  Here, we split the capacity
of each unique resource (link between two sites) across every sub-problem when
using resource splitting. We compare this to the regular \tn procedure:
partitioning the network into $k$ disjoint networks, with every link appearing
in a single sub-problem. We see that total flow remains roughly the same with
high $k$ when using resource splitting. On the other hand, without resource
splitting, the flow is up to 15$\times$ lower for high $k$. This result
highlights the importance of resource splitting for problems with resources
that \emph{have} to be used by certain clients for high utility.

\subsection{Alternatives to Random Partitioning}

\begin{figure}
    \center
    \includegraphics[width=0.8\columnwidth]{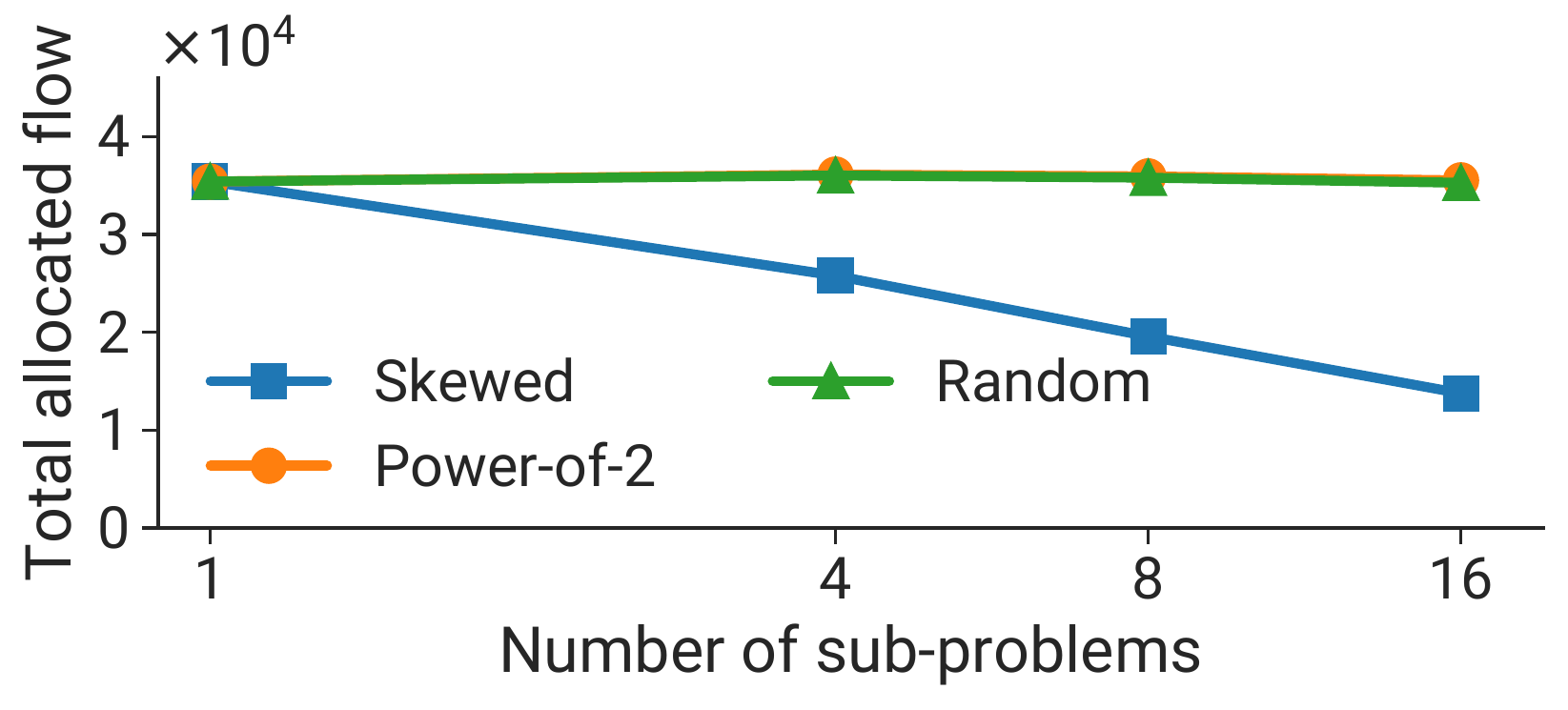}
    \vspace{-0.1in}
    \caption{
        Performance comparison of various partitioning algorithms for the
        max-flow objective in traffic engineering.
    }
    \label{fig:te_partitioning}
    \vspace{-0.1in}
\end{figure}

We implemented several algorithms to partition clients into sub-problems, and
compared them to random. Among these is a power-of-2 partitioning algorithm
that tries to assign each client sequentially to one of two randomly-chosen
sub-problems using ``distributional similarity'' to the original problem
as the metric.  We also implement a skewed partitioning algorithm that
deliberately creates skew among sub-problems to show the impact of bad
partitions.  Figure~\ref{fig:te_partitioning} shows the impact of these
partitioning algorithms on the quality of allocation returned by \tn on a
traffic engineering problem.  We see that random performs about as well as the
more sophisticated power-of-2 partitioning, while skewed partitions have poor
performance (skewed causes link congestion around certain nodes in the WAN).

%% file: tex/related_work.tex
\section{Related Work and Discussion}

In this section, we discuss other systems that use optimization problems to
allocate resources. We also comment on general efforts to accelerate solving
large optimization problems and how \tn fits into this body of work.

\paragraph{Optimization Problems in Systems.} A number of systems besides the
ones discussed in \S\ref{sec:prob-instances} use optimization problem
formulations to solve resource allocation problems.

TetriSched~\cite{tumanov2016tetrisched} is a cluster scheduler that is able to
leverage runtime predictions and deadline information (provided as input to the
system) to make smarter near-term decisions on how jobs should be allocated
resources, while also providing room for uncertainty from unknown future job
submissions. Preferences in resource space-time can be expressed in a new DSL
called STRL; these are then compiled down to a mixed-integer linear program
(MILP) whose solution describes when and how jobs should be executed.

RAS~\cite{ras} is a capacity reservation system that manages the allocation of
servers to clients within a datacenter region, while taking into account
failures, resource heterogeneity, and maintenance schedules. RAS formulates
problems as MILPs which are solved hourly.

DCM~\cite{suresh2020scalable} makes it easier to implement various cluster
management policies (e.g., ensure containers have enough of a particular
resource, or two containers are not placed in the same rack) by having users
specify cluster manager behavior declaratively through SQL queries written over
cluster state maintained in a relational database.  Similar to TetriSched,
these queries are then compiled down to an optimization problem that can be
solved by constraint solvers, such as CP-SAT~\cite{cpsat}. DCM supports
affinity and anti-affinity constraints.

Quincy~\cite{isard2009quincy} and Firmament~\cite{gog2016firmament} are
centralized datacenter schedulers that use efficient min-cost max-flow (MCMF)
based optimization to scale up to large clusters.

Another approach to quickly find good solutions is through variable
aggregation~\cite{litvinchev2013aggregation}, where a group of similar variables
is represented with a single meta-variable, and then an optimization problem
over the meta-variables is solved. The meta-solution can be used to derive a
solution to the original problem. NCFlow~\cite{abuzaid2021contracting} is such a
technique that solves the multi-commodity maximum flow problem (``total flow''
in \S\ref{sec:prob-te}). NCFlow divides a topology into geographic clusters,
and then solves a set of smaller-complexity flow problems; this yields faster
runtimes and fewer forwarding entries in the WAN topology, at the cost of a
smaller total flow. \Tn compares favorably to NCFlow, both in terms of solver
runtime and total flow (see Figure~\ref{fig:total_flow_and_runtime}).  While
\tn does not offer any reduction in forwarding entries, it is more general in
the objectives it can support; NCFlow only supports the max-flow objective. For
the max-flow objective, \tn and NCFlow can be used together to reduce solver
runtime and the number of forwarding entries.

SketchRefine~\cite{brucato2016scalablepackagequeries} uses a similar
approach to accelerate MILPs for ``packaging'' queries in a database, which
handle constraints and preferences over \emph{answer sets}. It uses a
quadtree-based partitioning algorithm to group tuples (rows in the relation)
into tuple subsets, and then uses an iterative reconciliation procedure to
convert initial per-group solutions into a global solution. SketchRefine's
partitioning step can be expensive (on the order of minutes), since it is
meant to be run over a fixed tuple set. While it is not clear how to
extend SketchRefine to resource allocation problems, which reason about
interactions between clients and resources, it offers another way to
quickly compute good solutions to certain types of large optimization
problems in systems.

\paragraph{Random Partitioning in Systems.} Random assignment has seen
success in other important systems problems as well. For example, in data
center networking \cite{singla2012jellyfish}, random graph topologies work
surprisingly well compared to commonly-used structured topologies such as
FAT-trees. In load-balancing algorithms~\cite{mitzenmacher2001power}, assigning
jobs to the least-loaded of just two randomly selected servers in a cluster
can drastically reduce the probability of overloading a server.

\paragraph{Approximation Algorithms.} FPTAS algorithms~\cite{fptas} return
results with a guaranteed approximation ratio and run in polynomial time over
this approximation factor. Proving an approximation ratio with \tn is hard
since we apply \tn to many different problems with various structures, as
opposed to designing a problem-specific approximation algorithm.

\paragraph{More Efficient Solving.} The optimization community has developed
various methods for scaling optimization solvers to handle large problems.
Fundamentally, these approaches rely strictly on identifying and then
exploiting certain mathematical structures (if they exist) within the problem
to extract parallelism; they make no domain-aware assumptions about the
underlying problem.  For example, Benders'
decomposition~\cite{geoffrion1972generalized, rahmaniani2017benders} only
applies to problems that exhibit a block-diagonal structure;
ADMM~\cite{boyd2011distributed, o2013splitting} has been applied to select
classes of convex problems, and Dantzig-Wolfe
decomposition~\cite{dantzig1960decomposition}, while more broadly applicable,
offers no speedup guarantee. This poses a significant limitation when applying
these methods to real-world systems, which often do not meet their criteria or
would need mathematical analysis to determine if this structure exists.

As mentioned in \S\ref{sec:primal_decomposition}, \tn can be interpreted as the
first iteration of primal decomposition for optimization problems with
separable objectives and certain types of coupled
constraints~\cite{boyd2007notes}.  By randomly partitioning large numbers of
clients and equally apportioning resources into sub-problems, we found that it
is possible to obtain high-quality solutions with a single iteration for a
broader set of allocation problem formulations, including MILPs.

%% file: tex/conclusion.tex
\section{Conclusion}

In this paper, we showed how a number of resource allocation problems in
computer systems are \emph{\apn} and proposed an efficient new method to solve
them. Such \apn allocation problems can be partitioned into more tractable
sub-problems by randomly assigning clients and resources.  Our technique, POP,
achieves strong results across a variety of tasks, including cluster
scheduling, traffic engineering, and load balancing, with runtime improvements
of up to 100$\times$ with small optimality gap, and outperforms greedy ad-hoc
heuristics.  We hope this work motivates using \tn as a simple pre-solving step
when solving optimization problems that arise in computer systems.

%% file: tex/acknowledgements.tex
\section*{Acknowledgements}

We thank our shepherd, Malte Schwarzkopf, the anonymous SOSP reviewers, Keshav
Santhanam, Kostis Kaffes, Akshay Narayan, Shoumik Palkar, and Deepti Raghavan
for their feedback that improved this work. We are also grateful to production
teams at Microsoft for facilitating access to datasets. This research was
supported in part by affiliate members and other supporters of the Stanford
DAWN project---Ant Financial, Facebook, Google, and VMware---as well as Toyota
Research Institute, Cisco, SAP, and the NSF under GRF grant DGE-1656518 and
CAREER grant CNS-1651570. Any opinions and conclusions expressed in this
material are those of the authors and do not reflect the views of the NSF.

%% file: tex/appendix.tex
\section*{Appendix (Not Peer-Reviewed)}

\section{Proof of Bound on Random Partitioning for Simple Allocation Problem}
\label{app:proof_body}

In this section, we show a full derivation of Equation~\ref{eq-perf_bound2},
which upper bounds the probability of a large gap between the optimal solution
and solution returned by \tn.

To quantify the gap between POP and optimal solutions, we need a sense of how
big $q_{s,t}$ -- the number of misplaced jobs of type $s$ in sub-problem $t$ --
is in practice. In this section, we assume that the number of resources of each
type are not necessarily equal; we define $n_s$ as the number of resources of
type $s$.  We can compute a probabilistic upper bound on $q_{s,t}$ using a
classical Chernoff bound, interpreting the random assignment of all type-$s$
jobs ($n_s$ of them) to sub-problems as Bernoulli trials where the probability
that any given type-$r$ job is placed in sub-problem $k$ is $1/k$.  Define
$X_{s,t}$ to be the sum of all such trials, i.e., the number of type-$s$ jobs
in sub-problem $t$, with $E[X_{s,t}] = n_s/k$.  Note that when $X_{s,t}$
exceeds the expected value, we get $X_{s,t} = n_s/k + q_{s,t}$. The Chernoff
upper bound~\cite{mitzenmacher2017probability} can then be used to find the
upper limit on the probability that the value of $X_{s,t}$ exceeds the expected
value by a fraction $\delta$:
\begin{align}
\Pr[X_{s,t} \geq (1+\delta)n_s/k] &= \Pr[q_{s,t} \geq \delta n_s/k] \nonumber \\
&\leq \exp \left(\frac{-\delta^2 n_s}{(2+\delta)k} \right) \label{eq-chernoff}
\end{align}

In the rest of this text, to simplify notation, we will refer to the RHS of
Equation~\ref{eq-chernoff} as $C(\delta,n_s, k)$.  For a simple problem with
$r=2$ and $k=2$, if we have $n=m=10^5$ jobs and resources split equally across
resource types, the probability of exceeding the expected amount of type $A$
jobs in a given sub-problem by 1\% is  0.2877, by 2\% is 0.00694, and by 3\% is
0.0000145.

This bound can be extended to misplaced jobs across all resource types and
sub-problems using the union bound, i.e., $\Pr(Z_1 \vee Z_2) \leq \Pr(Z_1) +
\Pr(Z_2)$. This can be used to compute an upper limit on the probability that
any resource type exceeds its expectation by a fraction ($1+\delta$) on any
sub-problem.  Define $Y_{s,t}$ to be the event that type-$r$ jobs in
sub-problem $k$ are in excess of the expected amount by a factor of
$(1+\delta)$, i.e., $X_{s,t} \geq (1+\delta)n_s/k$.  Then, we see that the
following holds:
\begin{align}
\Pr[Y_{s,1} \vee ... \vee Y_{s,k}] \leq \sum_{t = 1}^k \Pr[Y_t] \leq \sum_{t = 1}^k C(\delta,n_s, k)
\label{eq-partial_bound}
\end{align}

We can extend this to all resource types similarly.  Let $Z_{r}$ be the
probability that type-$s$ jobs in any sub-problem $k$ exceeds
$(1+\delta)n_s/k$. Using the union bound again, we can extend
Equation~\ref{eq-partial_bound} to compute the upper limit on the probability
that the total number of misplaced jobs exceeds $\delta n$.
\begin{align}
&\Pr\left[\sum_{s=1}^r \sum_{t=1}^k q_{s,t} \geq \delta n \right] \leq \Pr[Z_{1} \vee ... \vee Z_{R}] \nonumber \\
& \leq \sum_{s=1}^{r} \Pr[Z_{j}] \leq  \sum_{s = 1}^r \sum_{t = 1}^k C(\delta,n_s, k)\label{eq-allbound}
\end{align}

We can now combine this with Equation~\ref{eq-utility_bound} to bound the
performance of a randomized POP solution for the simplified allocation problem
discussed in \S\ref{sec:simple_partitioning_problem}. We define $\Gamma^*$ to
be an optimal allocation, $\Gamma^\text{POP}$ to be the allocation returned by
the \tn procedure, and $U(): \Gamma \rightarrow u$ to be a function that
computes the utility of an allocation $\Gamma$. Using
Equations~\ref{eq-utility_bound} and \ref{eq-allbound}, the probability that a
random job partition will result in a utility that is greater than $\delta
u_{\text{maxgap}}n$ from optimal is:
\begin{align}
\Pr[U(\Gamma^*) &- U(\Gamma^\text{POP}) \geq \delta u_{\text{maxgap}}n] \nonumber\\
&\leq  \Pr\left[\sum_{s = 1}^r \sum_{t = 1}^k q_{s,t}u_{\text{maxgap}} \geq \delta u_{\text{maxgap}}n\right] \nonumber \\
&\leq  \sum_{s = 1}^r \sum_{t = 1}^k C(\delta,n_s, k) \nonumber
\end{align}